\pdfoutput=1

\documentclass{raa}           

\usepackage{graphicx,times}
\usepackage{natbib}
\usepackage{amssymb,amsmath}
\usepackage{longtable, tabu}
\usepackage{tikz}
\usepackage{url}
\usetikzlibrary{shapes.geometric}
\usetikzlibrary{calc,shadows}
\bibpunct{(}{)}{;}{a}{}{,}

\usepackage[pagebackref=true,hidelinks]{hyperref}
\usepackage{multirow}
\newcommand*{\dif}{\mathop{}\!\mathrm{d}}

\begin{document}

   \title{A meta-analysis of molecular spectroscopy databases, and prospects of molecule detection with some future facilities}

 \volnopage{ {\bf 20XX} Vol.\ {\bf X} No. {\bf XX}, 000--000}
   \setcounter{page}{1}

   \author{Xin Liu\inst{1,2}, Fujun Du\inst{1,2}}

   \institute{ Purple Mountain Observatory, Chinese Academy of Sciences, Nanjing 210023, China; {\it fjdu@pmo.ac.cn}\\
        \and
             School of Astronomy and Space Science, University of Science and Technology of China, Hefei 230026, China\\
\vs \no
   {\small Received 20XX Month Day; accepted 20XX Month Day}
}

\abstract{Molecules reside broadly in the interstellar space and can be detected via spectroscopic observations. To date, more than 271 molecular species have been identified in interstellar medium or circumstellar envelopes. Molecular spectroscopic parameters measured in laboratory make the identification of new species and derivation of physical parameters possible. These spectroscopic parameters are systematically collected into databases, two of the most commonly used being the CDMS and JPL databases. While new spectroscopic parameters are continuously measured/calculated and added to those databases, at any point in time it is the existing spectroscopic data that ultimately limits what molecules can possibly be identified in astronomical data. In this work, we conduct a meta-analysis of the CDMS and JPL databases. We show the statistics of transition frequencies and their uncertainties in these two databases, and discuss the line confusion problem under certain physical environments. We then assess the prospects of detecting molecules in common ISM environments using a few facilities that are expected to be conducting spectroscopic observations in the future. Results show that CSST/HSTDM and SKA1-mid have the potential to detect some complex organic molecules, or even amino acids, with reasonable assumptions about ISM environments.
}

   \authorrunning{X. Liu \& F.J. Du }            
   \titlerunning{MS databases analysis and molecule detection with future facilities}  
   \maketitle

%
\section{Introduction}
\label{sec::intro}

The interstellar gases, including those in atomic, molecular, and ionized forms, account for $\sim$99\% of the mass of interstellar medium (ISM), with the rest being interstellar dust. Various molecules exist in the ISM, some of which become the tracers to study the formation of stars \citep{1987ARA&A..25...23S} and planets \citep{1976PThPh..56.1756A, 1980PThPh..64..544M}. Spectral lines of these molecules can be used to derive the gas temperature, velocity, column density, and internal dynamics. Therefore, interstellar molecular spectroscopy have become one of the most commonly used diagnostic tools for ISM.

The first interstellar molecule, CH, was reported to exist through observation of absorption feature in the optical band \citep{1940PASP...52..187M}. After the advent of radio astronomy, many more molecular species were detected thanks to the increasing sensitivity and spatial/spectral resolution of facilities. As of 2021, 241\footnote{At the time of writing, the number of interstellar molecules included in CDMS is 271 (https://cdms.astro.uni-koeln.de/classic/molecules).} \citep{2022ApJS..259...30M} molecular species consisting of 18 elements were reported to exist in ISM or circumstellar envelopes (CSE) (including unconfirmed ones). These molecules are made up of 2$-$70 atoms, and their transition frequencies range from microwave to ultraviolet bands \citep{2022ApJS..259...30M}.

The identification of interstellar molecules, in essence, is to infer the abundances of different molecules and physical parameters of interstellar gas. Technically, this is an optimization problem. It tries to find the best set of parameters with regard to some optimization goal, which is usually the difference between the simulated and observed spectrum. In the case of a positive identification, the individual lines in simulated spectrum is supposed to match those in observation with the right line widths and intensities \citep{2008A&A...482..179B}. The simulated spectrum is calculated using the radiative transfer equation and usually (but not always) under the assumption of local thermodynamic equilibrium (LTE). As key parameters in the radiative transfer equation, molecular spectroscopy data are the foundation of the molecule identification process.

Databases such as the one maintained by Jet Propulsion Laboratory\footnote{\url{https://spec.jpl.nasa.gov/ftp/pub/catalog/catdir.html}} (JPL), the Cologne Database for Molecular Spectroscopy\footnote{\url{https://cdms.astro.uni-koeln.de/classic/entries/}} (CDMS), the HIgh-resolution TRANsmission molecular absorption database\footnote{\url{https://hitran.org}} (HITRAN), and the NIST Atomic Spectra Database \footnote{\url{https://physics.nist.gov/PhysRefData/ASD}} provide spectroscopy data to simulate spectral lines. Among these databases, JPL and CDMS are frequently used by radio astronomers in molecular identification. In this paper we present a meta-analysis of the JPL and CDMS spectroscopic databases.

A meta-analysis of these two databases can tell us which species have the potential to be detected at present and which transition lines are more likely to be resolved under certain astrophysical environments. We can estimate the integration time needed for a certain facility to detect the spectral lines of any molecular species with given astrophysical parameters. In general, a meta-analysis of current molecular databases may provide useful insights for future molecular identification. In Section \ref{sec::Iden_Proc} we present a brief description of radiative transfer, followed up with the meta-analysis of molecular spectroscopy databases in Section \ref{sec::Databases}. We discuss what molecules can be observed and how much time it will cost to observe certain molecular species in Section \ref{sec::Instruction}, followed by a conclusion in Section \ref{sec::conclusion}.

\section{Radiative transfer}
\label{sec::Iden_Proc}
About 90 percent molecules are detected in radio frequency range \citep{2022ApJS..259...30M}. This is mainly because of the composition of molecules and the environment where they reside in: the low temperature of molecular clouds means that the electronic and vibrational emissions of molecules cannot occur (absorption of background sources do sometimes occur); the large moment of inertia of these molecules (except for molecular hydrogen) lead to small rotational constants thus low transition frequencies.

The interactions of radiation field with matter can be expressed by the radiative transfer equation \citep{1979rpa..book.....R} as
\begin{equation}
	\frac{\dif I_{\nu}}{\dif s} = -\alpha_{\nu}I_{\nu} + j_{\nu},
\end{equation}
where $\alpha_{\nu}$ and $j_{\nu}$ are the absorption and emission coefficient, respectively, $I_{\nu}$ the specific intensity, $s$ the travel path. This equation can be integrated as
\begin{equation}
	\label{equ::radiative_e}
	I(\nu) = I_{0}(\nu)e^{-\tau_{\nu}} + S_{\nu}(T)(1-e^{-\tau_{\nu}}).
\end{equation}
Here $S_{\nu}$ is referred to as the source function, which in LTE assumption equals the Planck blackbody radiation function $B_{\nu}$ of the same temperature. $ I_{0}(\nu)$ is the background radiation, $T$ the excitation temperature, and $\tau$ the optical depth which is defined as the integration of $\alpha_{\nu}$ along the travel path $s$:
\begin{equation}
	\tau_{\nu} = \int \alpha_{\nu} \dif s = \frac{c^{2}}{8\pi \nu_{0}^{2}} \frac{g_{u} e^{-E_{l}/kT}}{Q(T)} N_{\mathrm{tot}} A_{ul} [1 - e^{-h\nu_{0}/kT}]\varphi(\nu),
\end{equation}
where $T$ is assumed to be independent of $s$, $c$ the speed of light, $\nu_{0}$ the transition frequency, $g_u$ the statistical weight of the $u$ energy level, $E_l$ the energy of $l$ level, $Q(T)$ the partition function, $N_{\mathrm{tot}}$ the total column density, $A_{ul}$ the Einstein $A$ coefficient, $h$ the Planck constant, $k$ the Boltzmann constant, and $\varphi(\nu)$ the line profile function. The Einstein $A_{ul}$ coefficient describes the probability per unit time of a molecule in an excited energy level $E_{u}$ to return spontaneously to a lower energy level $E_{l}$ \citep{2009tra..book.....W}. $A_{ul}$ is related with the other two Einstein coefficients $B_{ul}$ and $B_{lu}$ via
\begin{equation}
	\label{equ::Aul}
	A_{ul}=\frac{8\pi h\nu_{0}^{3}}{c^{3}}B_{ul},
\end{equation}
\begin{equation}
	\label{equ::Bul}
	g_{l}B_{lu}=g_{u}B_{ul}.
\end{equation}
in which $g_l$ is the statistical weight of the $l$ energy level. Molecular spectroscopy data, including $Q(T)$, $g_{u}$, $E_{l}$, $A_{ul}$, and $\nu_{0}$, are needed when calculating the spectral line intensity, which can be accessed through several databases.

\section{A Meta-analysis of Molecular Spectroscopy Databases}
\label{sec::Databases}

\subsection{The CDMS and JPL spectroscopic databases}
The JPL provides spectroscopy data ``based on the project needs of astronomers" \citep{1998JQSRT..60..883P} and CDMS catalogs ``molecules that are present or are supposed to be present in ISM or CSE or in planetary atmospheres" \citep{muller_cologne_2001}. CDMS/JPL provides spectroscopic data for 1137/401 species, of which 189 are available in both databases. Each line of a data file contains the frequency $\nu$ and its uncertainty, the base 10 logarithm of integrated intensity $I$ at 300\,K, degree of freedom in the rotational partition function, lower state energy $E_{l}$, upper state degeneracy $g_{u}$, and the quantum numbers. These data can be accessed through their websites and downloaded in ASCII format. The Einstein $A$ coefficient can be calculated as \citep{1998JQSRT..60..883P}:
\begin{equation}
	A_{ul}=I(300\,K)\nu^{2}\frac{Q(T)}{g_{u}(e^{-E_{l}/kT}-e^{-E_{u}/kT})}\times2.7964\times10^{-16}
\end{equation}
where $E_{u}$ is the upper state energy. CDMS provides an interface to search for all transitions within a given frequency range, and plot their intensities as a function of frequency under some simple assumptions online \citep{muller_cologne_2005}.

Regarding the 189 species that are included in both the CDMS and JPL databases, their entries can be classified into four cases. Some of them contain completely identical transitions except for the precision of spectroscopy data. The transitions of several species in one database are the subsets of those in another. Some pairs have a part of identical transitions (again, except for the precision) and additional unique ones in each database. The rest pairs are complementary to each other, with no repeating transition.

CDMS also participates in the Virtual Atomic and Molecular Data Centre and provides data in MySQL database format\footnote{\url{http://cdms.ph1.uni-koeln.de/cdms/portal/cdms\_lite.db.gz}} \citep{endres_cologne_2016}. Different from the files directly downloaded from their website, this database file contains the Einstein $A$ coefficients rather than the integrated intensity. Molecular species in JPL are also included in this database file.

\subsection{Transition Frequency Distribution}
\label{sec::Tran_Freq_Distr}

\begin{figure}[!htb]
	\centering
	\includegraphics[width=10cm]{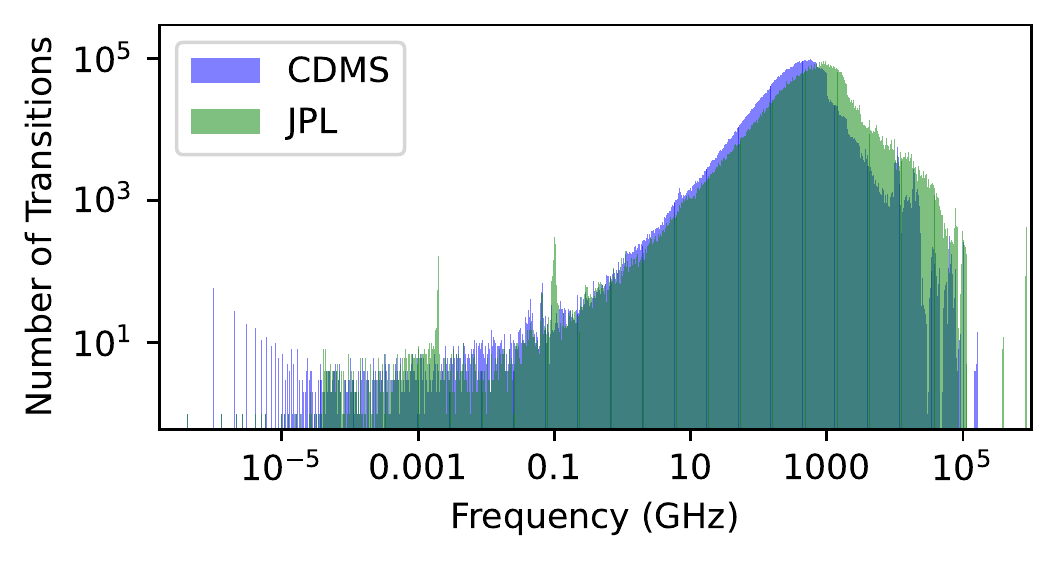}
	\caption{Histograms of the transition frequencies of molecules in CDMS and JPL. Each bin represents the velocity width of 10,000\,km/s. Data from CDMS  are in purple and JPL in light green. }
	\label{fig::trans_distr_AS_All}
\end{figure}

The transition frequencies range from 0.4\,kHz to $\sim$166,000\,GHz in CDMS and 0.4\,kHz to $\sim$857845\,GHz in JPL. In Fig. \ref{fig::trans_distr_AS_All} we show the histogram of transition frequencies of all the species in CDMS and JPL in a log$-$log scale. The number of transitions in each 10,000\,km/s velocity interval rises from $\sim$10 at $\sim$0.01\,MHz to $\sim$100,000 at $\sim$1000\,GHz and then decreases sharply. Quantitatively, about 99\%, 90\%, and 50\% of these frequencies fall in the intervals from 8 to 21,420\,GHz, 70 to 3098\,GHz, and 266 to 985\,GHz, respectively. 
Nearly two hundred of transitions have extremely low frequencies ($<10^{-5}$\,GHz), which are all from the rotational transitions of CH$_3$OH, CH$_3^{18}$OH, and H$_2$CO. These three asymmetric top molecules are close to symmetric top structures, so the extremely low energy difference between two energy levels, which is caused by the deviation from the symmetric top case, leads to extremely low transition frequencies. They are calculated in theory but may not be detectable in practice because of the extremely low frequency and intensity.

\subsection{Frequency Uncertainty}

\begin{figure}[!htb]
	\centering
	\includegraphics[width=7cm]{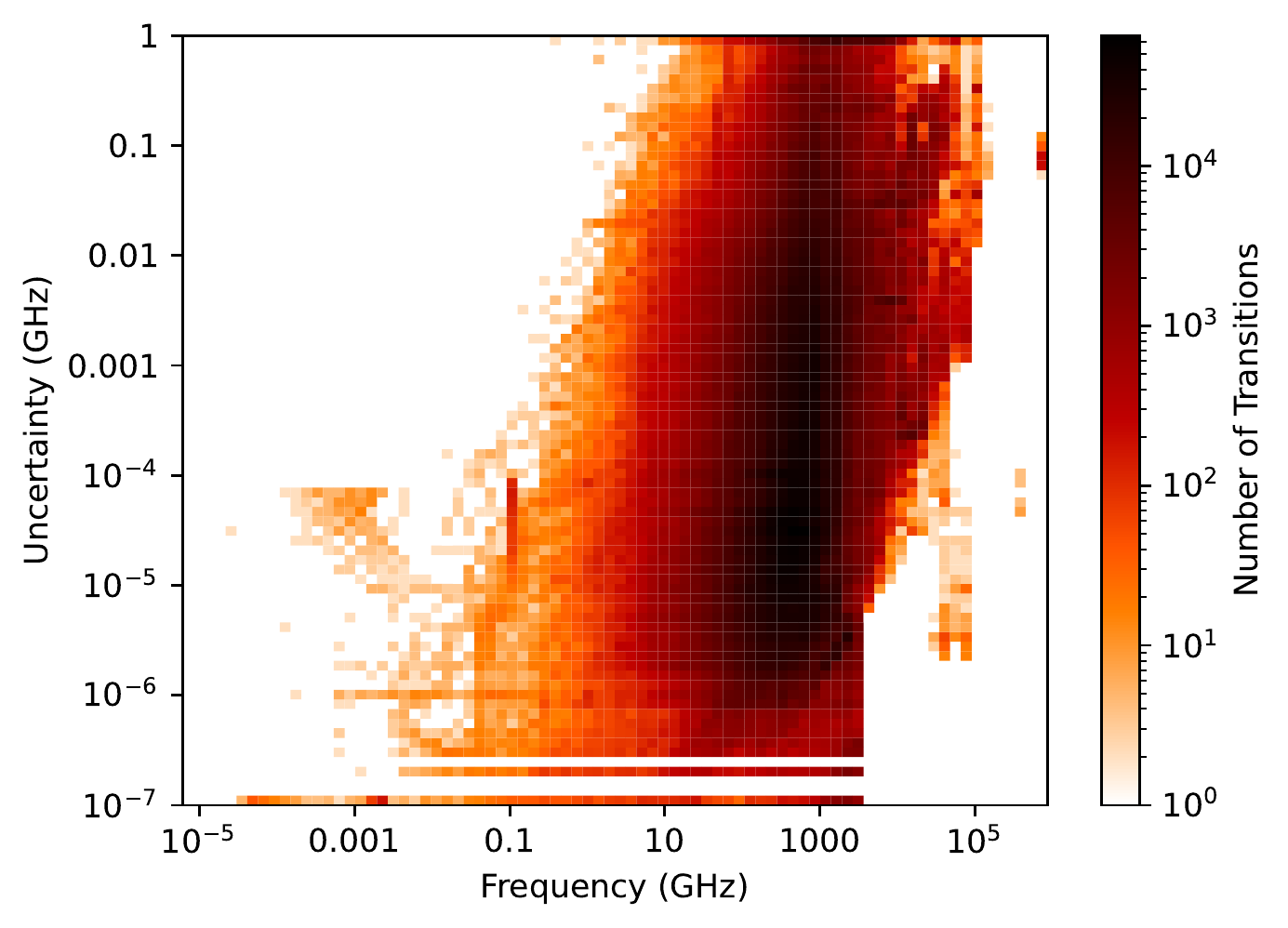}
	\includegraphics[width=7cm]{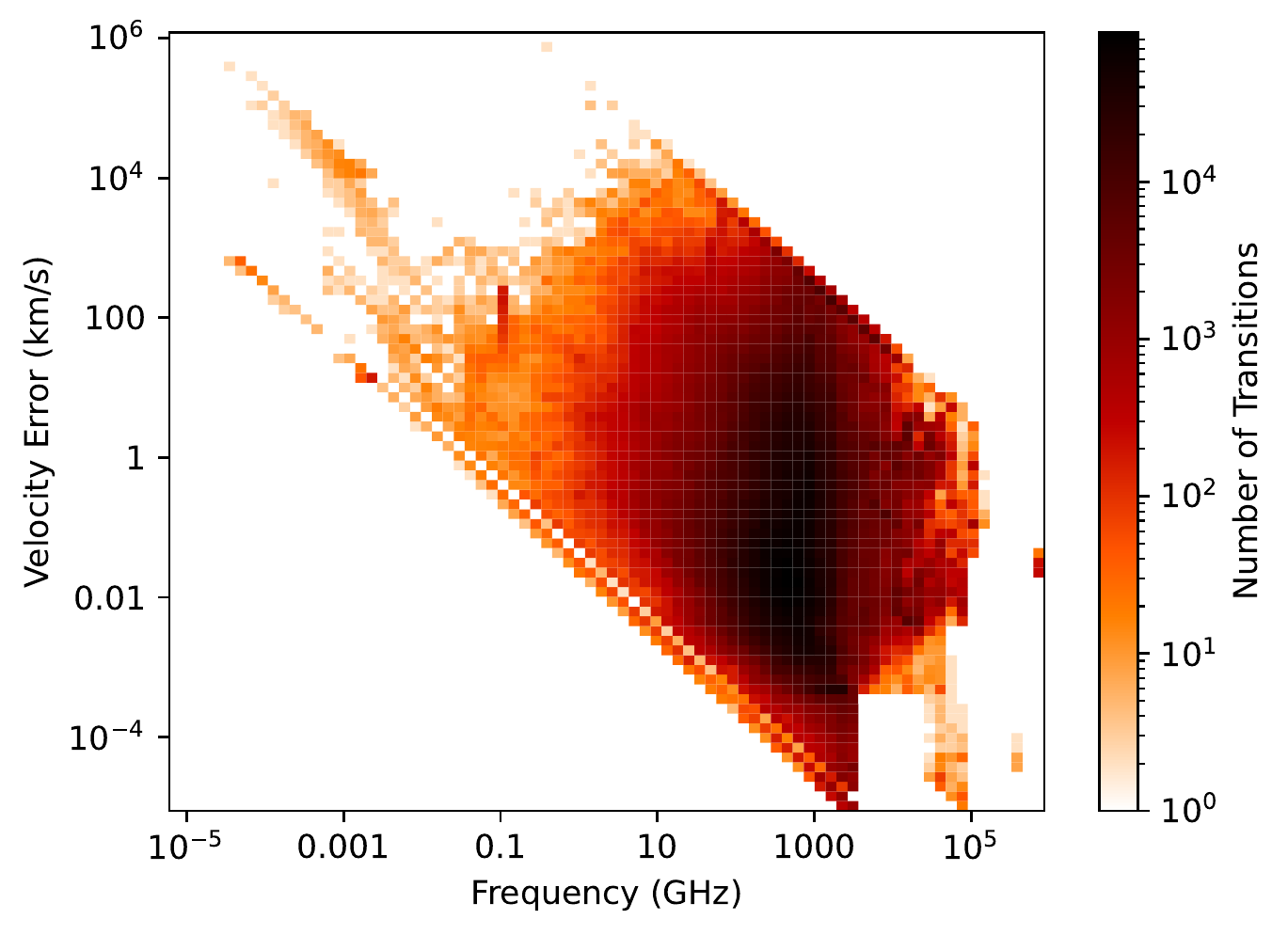}	
	\caption{2D histogram of frequency uncertainties and the corresponding velocity errors in both CDMS and JPL.}
	\label{fig::uncer_distr}
\end{figure}

Molecular spectroscopy data are either calculated theoretically or measured in laboratory, and the precision of these data determines the reliability of molecular identification. Files provided by CDMS and JPL include the uncertainty of transition frequencies, which ranges from 0$-$1\,GHz. This frequency error can be translated into a velocity error as
\begin{equation}
	v_{_{\mathrm{err}}} = \frac{\nu_{_{\mathrm{err}}}}{\nu}\cdot c,
\end{equation}
where $\nu$ is the transition frequency and $\nu_{\mathrm{err}}$ represents its uncertainty. Figure \ref{fig::uncer_distr} gives the distribution of frequency uncertainties and velocity errors of all transitions in both CDMS and JPL. Most of these uncertainties are low enough that the majority of velocity errors are under 1\,km/s.

\begin{figure}[!htb]
	\centering
	\includegraphics[width=12cm]{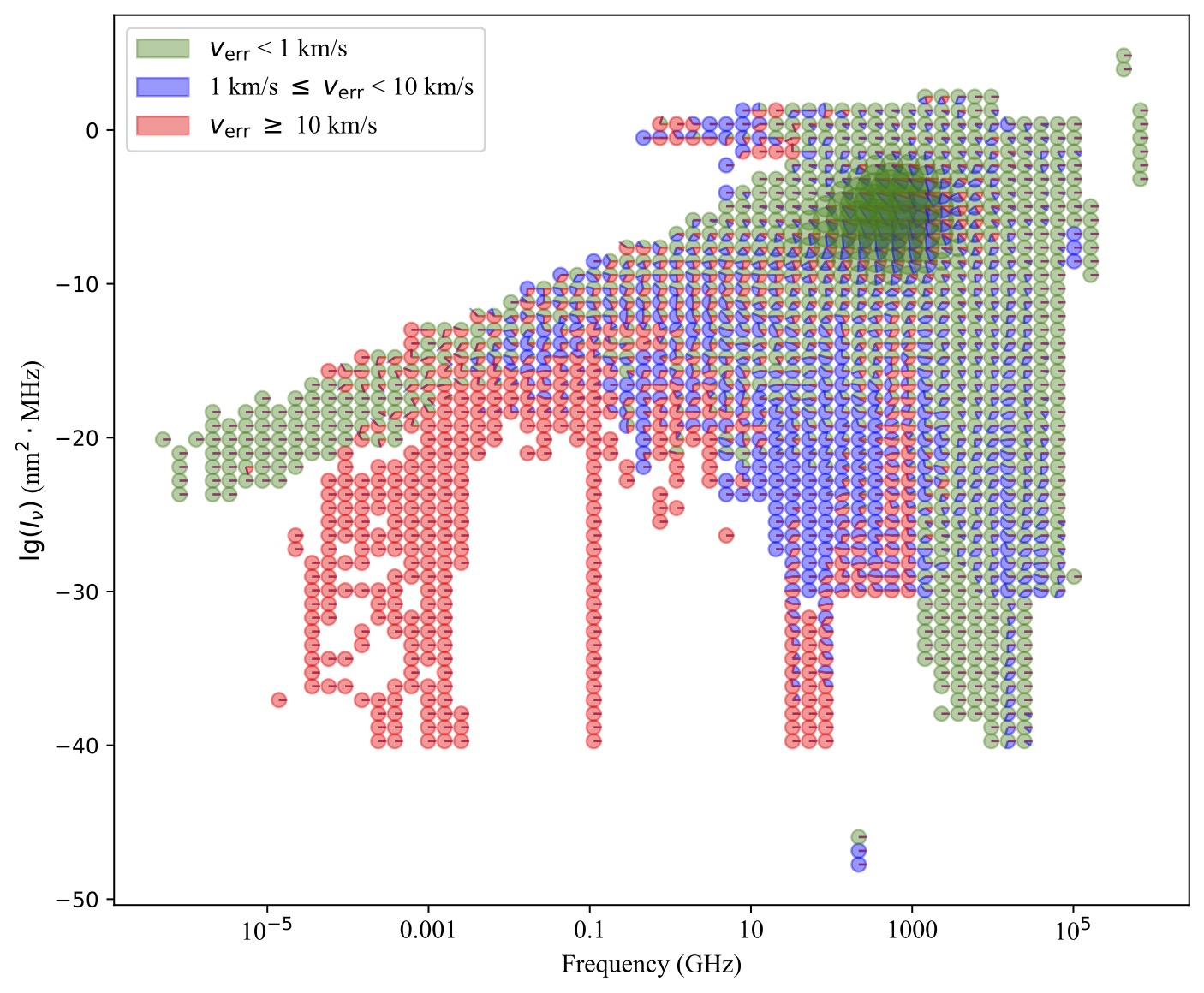}
	\caption{The distribution of velocity errors in a frequency-intensity map. Each pie chart was filled with at most three colors: green, purple, and red, corresponding to fractions of frequencies that are reliable, tentative, and unreliable, respectively. A bigger pie chart means a greater number of transition lines are cataloged in this frequency-intensity range.}
	\label{fig::Uncers_AS_All}
\end{figure}

The internal turbulent motion of gas flows make their radial velocities have a finite velocity width. Assuming a typical velocity width $\Delta v \lesssim$ 10\,km/s (FWHM), if the velocity error of a transition is comparable to or greater than the velocity width, it is difficult to attribute the spectral line to its corresponding molecule. For this reasons, we divided all velocity errors into three levels, $v_{{\mathrm{err}}}<$ 1\,km/s, 1\,km/s $\le v_{{\mathrm{err}}}<$ 10\,km/s, and $v_{{\mathrm{err}}}\ge$ 10\,km/s, as reliable, suspicious, and unreliable, respectively. We assigned the three levels with different colors and drew a velocity error distribution map in Fig. \ref{fig::Uncers_AS_All}, 
where we can see that except for those calculated theoretically (uncertainty$=0$\,MHz), spectroscopy data with frequencies lower than 0.1\,GHz are less reliable than those with higher frequencies, and data with higher intensities are also more reliable. All data with relatively low frequencies ($\nu < $0.1 \,GHz) and low intensity ($\lg I_{\nu} < -25$\,$\mathrm{nm^{2}}\cdot \mathrm{MHz}$) have high velocity errors ($v_{{\mathrm{err}}}>$ 10\,km/s), and they are all from the isotopomers of either O$_{2}$ or OH. The relatively large pie charts are in the 100$-$1000\,GHz range, which means more transitions are cataloged in this frequency range. In general, most of these transitions are reliable.

\subsection{Line Confusion}
\label{sec::line_confusion}
If two transition lines have similar intensities, and the separation of line centers is less than either of their line widths, these two lines are considered to be blended, which is known as the line confusion problem. To discuss this issue, we assume that all previously identified molecules are present in an interstellar cloud, and have a constant excitation temperature of 20\,K and an FWHM velocity width of 5\,km/s. The column densities are set to their typical values that obtained from literature\footnote{These values and corresponding references are available on Github: \url{https://github.com/LiuXnhhh/paper_materials}}. Those without definite literature values are all belong to vibrational/electronic excited states or minor isotopomers. For the latter ones we adopt empirical isotope ratios of $\text{H}/\text{D}=1000$ \citep{2013A&A...549L...3P}, $^{12}\text{C}/^{13}\text{C}=59$,  $^{14}\text{N}/^{15}\text{N}=237$, $^{32}\text{S}/^{34}\text{S}=19$ \citep{1998A&A...337..246L}, $^{16}\text{O}/^{18}\text{O}=672$ \citep{1981ApJ...243L..47W}, $^{16}\text{O}/^{17}\text{O}=1935$ \citep{2008A&A...487..237W}, $^{28}\text{Si}/^{29}\text{Si}=13$, $^{28}\text{Si}/^{30}\text{Si}=12$ \citep{1980ApJ...242.1005W}, and $^{35}\text{Cl}/^{37}\text{Cl}=2.1$ \citep{2012ApJ...751L..37D} to derive rough column density values from their main isotopomer instead, while those in vibrational/electronic excited states are not included for the lack of good estimations of their column densities. We must note that the H/D ratio varies dramatically under the influence of difference chemical processes in low temperature, and a reasonable mean value of 1000 is adopted here. We then define the line confusion as the number of spectral lines in each 10\,km/s interval in the simulated spectrum, and arbitrarily call intervals with more than five transitions as ``crowded". Lines with extremely low intensities ($T_{\mathrm{mb}}<0.01$\,K) are not included for line-confusion counting.

\begin{figure}[!htb]
	\centering
	\includegraphics[width=12cm]{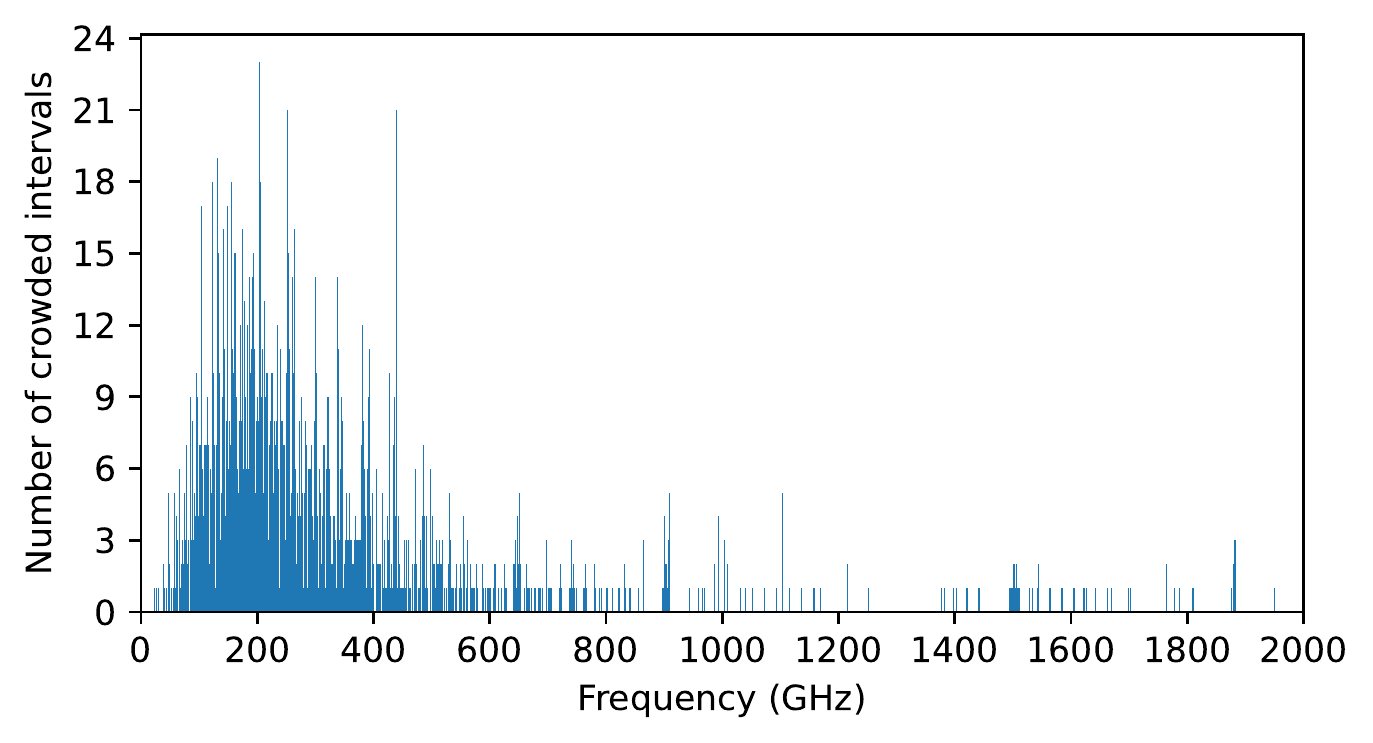}
	\caption{The distribution of ``crowded" intervals which is defined in Section \ref{sec::line_confusion}. Each bin represents a frequency width of 1\,GHz.}
	\label{fig::Overlapped_trans}
\end{figure}

Figure \ref{fig::Overlapped_trans} shows the number of crowded intervals in each 1\,GHz bin from 0 to 2000\,GHz,
from which we know that transitions between $\sim$50 and 600\,GHz are more crowded than those in higher or low frequency ranges. Note that all previously detected molecules are included to give an overview of the line confusion, but this may oversimplify the reality. A source is unlikely to have all these molecules, and the excitation conditions of molecules may differ from one to another. Some sources contain multiple velocity components, which may aggravate the problem of line confusion.

\section{Suggestions for Future Observations}
\label{sec::Instruction}

For a given source, we would like to know which transitions can be obtained at certain noise level using a given telescope within a reasonable integration time, and we also want to figure out how much time it would take if we intend to search for certain molecular species, or what physical environment should this molecule reside in if it is detectable.

With regard to these concerns, we investigate the observability of molecules using a few major facilities that would be operating in the future. To calculate the line intensities, we consider three types of ISM environments: i) a cold diffuse source with an angular size ($\theta_{\text{s}}$) of 60\,$''$ and an excitation temperature ($T_{\mathrm{ex}}$) of 10\,K \citep{1997ApJ...486..862P}; ii) a hot dense core with an angular size of 3\,$''$ and an excitation temperature of 200\,K \citep{2013A&A...559A..47B}; iii) an intermediate case with an angular size of 40\,$''$ and an excitation temperature of 30\,K. These three cases are summarized in Table \ref{tab::common_environment}.
\begin{table}[!htb]\small
	\begin{center}
		\caption[]{Three sets of assumed physical parameters for simulating molecular spectra.}\label{tab::common_environment}
		\begin{tabular}[5pt]{ccc}
			\hline\noalign{\smallskip}
			&  $T_{\mathrm{ex}}$ (K) & $\theta_{\text{s}}$ ($''$)  \\
			\hline\noalign{\smallskip}
			Case 1  & 10 & 60 \\
			Case 2 & 200 & 3 \\
			Case 3 & 30 & 40 \\
			\noalign{\smallskip}\hline
		\end{tabular}
	\end{center}
\end{table}
Column densities are set to their typical measured values, as mentioned in Section \ref{sec::line_confusion}. A typical line width of 5\,km/s is assumed for all lines. The beam filling factor for a source $s$ is defined as
\begin{equation}
	\eta_{\text{s}} = \frac{\theta_{\text{s}}^{2}}{\theta_{\text{s}}^{2}+\theta_{\text{t}}^{2}},
\end{equation}
where $\theta_{\text{t}}$ is the half-power beamwidth of a telescope. This factor becomes important when the source is not extended. The cosmic microwave background (CMB) radiation ($T_{_{\mathrm{CMB}}}=2.725$\,K) is also considered. Lines with the peak intensities higher than three times sensitivity ($\sigma$) are taken as detectable, and molecular species with more than three detectable transitions are regarded as unambiguous identifications \citep{2005ApJ...619..914S}. Molecular spectroscopy data used in this section are mainly taken from CDMS and is supplemented by those in JPL. Note that the column densities of nearly one third of molecules are derived from empirical isotope ratios. These molecules are either minor isotopomers or in a vibrational/electronic excited state thus may be less abundant than assumed, causing the overestimation of their line intensities. Thus the detectabilities stated here can only be viewed as an optimistic estimation, and is valid only for molecules under LTE conditions. To ensure the reliability of the predictions, we have compared the synthesized spectra with the line survey reported by \citet{2013A&A...559A..47B} using the same physical parameters. Results show that our simulated lines are almost identical to their synthetic spectra. We will benchmark the performance of current spectral line modeling programs and discuss this in future.

\subsection{The Terahertz module of CSST}

The high-sensitivity terahertz detection module (HSTDM) \citep{8563600} onboard Chinese Space Station Telescope (CSST) covers a frequency range of 410$-$510\,GHz. CSST has a diameter of 2\,m, which means at the working band of HSTDM the beamwidth is $\sim$70\,$''$. With a frequency resolution of 0.1 MHz, the receiver promises to be able to reach a $3\sigma$ noise level of 0.15\,K with 200\,s of integration.

\subsubsection{Molecules that are observable in one hour}
Among all previously identified molecules and their isotopomers, 25/21/58 of them can be detected by CSST/HSTDM within one hour in case 1/2/3, respectively (see Tables \ref{tab::all_detectable_CSST_1}, \ref{tab::all_detectable_CSST_2}, and \ref{tab::all_detectable_CSST_3}). We take two molecules of special significance or with relatively more detectable transitions as examples and discuss their observability in detail.

\begin{table}[!htb]\tiny
	\begin{center}
		\caption[]{List of potentially detectable molecules with one hour of integration by CSST/HSTDM in case 1.}\label{tab::all_detectable_CSST_1}
		\begin{tabular}[5pt]{lllllll}
			\hline\noalign{\smallskip}
			2 Atoms &  3 Atoms  & 4 Atoms & 5 Atoms & 6 Atoms& 8 Atoms & 9 Atoms  \\
			\hline\noalign{\smallskip}
			$\mathrm{CN}$	 & $\mathrm{C_{2}H}$	 & $\mathrm{DNCO}$	 & $\mathrm{CH_{2}NH}$	 & $\mathrm{^{13}\!CH_{3}OH}$	 & $\mathrm{H_{2}NCONH_{2}}$	 & $\mathrm{CH_{3}CH_{2}CN}$	\\
			$\mathrm{NO}$	 & $\mathrm{DCO^{+}}$	 & $\mathrm{H_{2}CO}$	 & $\mathrm{HCOCN}$	 & $\mathrm{CH_{3}OH}$	 & 	 & $\mathrm{^{13}\!CH_{3}CH_{2}CN}$	\\
			& $\mathrm{H_{2}Cl^{+}}$	 & $\mathrm{HONO}$	 & $\mathrm{HNCNH}$	 & $\mathrm{H_{2}CCNH}$	 & 	 & $\mathrm{CH_{3}^{13}\!CH_{2}CN}$	\\
			& $\mathrm{NH_{2}}$	 & $\mathrm{NH_{2}D}$	 & 	 & 	 & 	 & $\mathrm{C_{2}H_{5}^{13}\!CN}$	\\
			& $\mathrm{SO_{2}}$	 & 	 & 	 & 	 & 	 & $\mathrm{CH_{3}OCH_{3}}$	\\
			& $\mathrm{Si_{2}C}$	 & 	 & 	 & 	 & 	 & 	\\
			& $\mathrm{TiO_{2}}$	 & 	 & 	 & 	 & 	 & 	\\
			\noalign{\smallskip}\hline
		\end{tabular}
	\end{center}
\end{table}
\begin{table}[!htb]\tiny
	\begin{center}
		\caption[]{List of potentially detectable molecules with one hour of integration by CSST/HSTDM in case 2.}
		\label{tab::all_detectable_CSST_2}
		\begin{tabular}[5pt]{lllllll}
			\hline\noalign{\smallskip}
			2 Atoms &  3 Atoms  & 5 Atoms & 6 Atoms  & 8 Atoms & 9 Atoms & $>$  9 Atoms \\
			\hline\noalign{\smallskip}
			$\mathrm{AlO}$	 & $\mathrm{AlOH}$	 & $\mathrm{HC_{3}N}$	 & $\mathrm{^{13}\!CH_{3}OH}$	 & $\mathrm{H_{2}NCONH_{2}}$	 & $\mathrm{CH_{3}CH_{2}CN}$	 & $\mathrm{CH_{3}OCH_{2}OH}$ \\
			$\mathrm{CN}$	 & $\mathrm{DCO^{+}}$	 & 	 & $\mathrm{CH_{3}OH}$	 & $\mathrm{NH_{2}CH_{2}CN}$	 & $\mathrm{^{13}\!CH_{3}CH_{2}CN}$	 & $i$-$\mathrm{C_{3}H_{7}CN}$ \\
			$\mathrm{PO}$	 & $\mathrm{SO_{2}}$	 & 	 & $\mathrm{H_{2}CCNH}$	 & 	 & $\mathrm{CH_{3}^{13}\!CH_{2}CN}$	 &  \\
			$\mathrm{TiO}$	 & $\mathrm{TiO_{2}}$	 & 	 & 	 & 	 & $\mathrm{C_{2}H_{5}^{13}\!CN}$	 &  \\
			&	 & 	 & 	 & 	 & $\mathrm{C_{2}H_{5}C^{15}\!N}$	 &  \\
			\noalign{\smallskip}\hline
		\end{tabular}
		\vspace*{-1.5ex}
		\tablecomments{0.86\textwidth}{The descriptor ``$i$-" ($iso$) of $\mathrm{C_{3}H_{7}CN}$ means the four C atoms form a branched skeleton.}
	\end{center}
\end{table}
\begin{table}[!htb]\tiny
	\begin{center}
		\caption[]{List of potentially detectable molecules with one hour of integration by CSST/HSTDM in case 3.}\label{tab::all_detectable_CSST_3}		
		\begin{tabular}[5pt]{lllllllll}
			\hline\noalign{\smallskip}
			2 Atoms &  3 Atoms & 4 Atoms & 5 Atoms & 6 Atoms & 7 Atoms& 8 Atoms  & 9 Atoms & $>$  9 Atoms \\
			\hline\noalign{\smallskip}
			$\mathrm{AlF}$	 & $\mathrm{AlOH}$	 & $\mathrm{DNCO}$	 & $\mathrm{CH_{3}Cl}$	 & $\mathrm{^{13}\!CH_{3}OH}$	 & $c$-$\mathrm{C_{2}H_{4}O}$	 & $\mathrm{CH_{2}OHCHO}$	 & $\mathrm{^{13}\!CH_{3}^{13}\!CH_{2}CN}$	 & $\mathrm{CH_{3}CH_{2}OCHO}$ \\
			$\mathrm{AlO}$	 & $\mathrm{C_{2}H}$	 & $\mathrm{H_{2}^{13}\!CO}$ $\!^{*}$	 & $\mathrm{CH_{3}^{37}\!Cl}$	 & $\mathrm{CH_{3}OH}$	 & $s$-$\mathrm{H_{2}CCHOH}$	 & $\mathrm{H_{2}NCONH_{2}}$	 & $\mathrm{^{13}\!CH_{3}CH_{2}^{13}\!CN}$	 & $\mathrm{CH_{3}OCH_{2}OH}$ \\
			$\mathrm{^{13}\!CN}$	 & $\mathrm{DCO^{+}}$	 & $\mathrm{H_{2}CO}$	 & $\mathrm{H_{2}CNH}$	 & $\mathrm{H_{2}CCNH}$	 & 	 & $\mathrm{NH_{2}CH_{2}CN}$	 & $\mathrm{^{13}\!CH_{3}CH_{2}CN}$	& $aGg'$-$\mathrm{(CH_{2}OH)_{2}}$ \\
			$\mathrm{CN}$	 & $\mathrm{H_{2}Cl^{+}}$	 & $\mathrm{HNCO}$	 & $\mathrm{HCOCN}$	 & 	 & 	 & 	 & $\mathrm{C_{2}H_{5}^{13}\!CN}$	& $gGg'$-$\mathrm{(CH_{2}OH)_{2}}$ \\
			$\mathrm{FeO}$	 & $\mathrm{HNO}$	 & $\mathrm{HONO}$	 & $\mathrm{HNCNH}$	 & 	 & 	 & 	 & $\mathrm{C_{2}H_{5}C^{15}\!N}$	 & $i$-$\mathrm{C_{3}H_{7}CN}$ \\
			$\mathrm{NO}$	 & $\mathrm{NH_{2}}$	 & $\mathrm{NH_{2}D}$	 & 	 & 	 & 	 & 	 & $\mathrm{CH_{2}DCH_{2}CN}$ $\!^{*}$	 & $n$-$\mathrm{C_{3}H_{7}CN}$ \\
			$\mathrm{NS}$	 & $\mathrm{SO_{2}}$	 & 	 & 	 & 	 & 	 & 	 & $\mathrm{CH_{3}^{13}\!CH_{2}^{13}\!CN}$	 &  \\
			$\mathrm{PO}$	 & $\mathrm{Si_{2}C}$	 & 	 & 	 & 	 & 	 & 	 & $\mathrm{CH_{3}^{13}\!CH_{2}CN}$	 &  \\
			$\mathrm{SO}$	 & $\mathrm{TiO_{2}}$	 & 	 & 	 & 	 & 	 & 	 & $\mathrm{CH_{3}CH_{2}CN}$	 &  \\
			$\mathrm{SiC}$	 & 	 & 	 & 	 & 	 & 	 & 	 & $\mathrm{CH_{3}CH_{2}OH}$	 &  \\
			$\mathrm{SiN}$	 & 	 & 	 & 	 & 	 & 	 & 	 & $\mathrm{CH_{3}CHDCN}$ $\!^{*}$	 &  \\
			$\mathrm{TiO}$	 & 	 & 	 & 	 & 	 & 	 & 	 & $\mathrm{CH_{3}OCH_{3}}$	 &  \\
			\noalign{\smallskip}\hline
		\end{tabular}
		\vspace*{-1.5ex}
		\tablecomments{0.86\textwidth}{$^{*}$Molecules with column densities derived from their main isotopomer. The descriptor ``$c$-" in $c$-$\mathrm{C_{2}H_{4}O}$ represents the O atom and two C atoms that are arranged in a $cyclic$ form. The ``$s$-" in $s$-$\mathrm{H_{2}CCHOH}$ means the ``$syn$-" form of the the molecule, with the H atom in hydroxyl and CH$_2$ ridical on the same side of the C$-$O bond. The ``$aGg'$-" and ``$gGg'$-" represent two rotational isomers of $\mathrm{(CH_{2}OH)_{2}}$: $G$ ($gauche$) means two hydroxymethyls are $\pm$60$^{\circ}$ apart, and $a$ ($anti$), $g$, $g'$ mean the hydroxyl and hydroxymethyl are 180$^{\circ}$ , 60$^{\circ}$, and -60$^{\circ}$ apart, repsectively. The ``$n$-" ($normal$) means the four C atoms in $\mathrm{C_{3}H_{7}CN}$ are arranged in a straight chain.}
	\end{center}
\end{table}

\paragraph{CH$_3$OH and $^{13}$CH$_3$OH} In case 1/2/3, 48/60/135 transition lines of CH$_3$OH and 98/322/265 lines of $^{13}$CH$_3$OH can be detected within one hour. As one of the most prominent molecules expected to be detected by CSST/HSTDM, several transitions of $^{13}$CH$_3$OH in 234$-$238\,GHz \citep{1985ApJS...58..341S} and 488 to 571\,GHz \citep{2007A&A...476..807P} have already been detected in different lines of sight. Interstellar CH$_3$OH, the main isotopomer of $^{13}$CH$_3$OH, can be used to trace the interface between outflows and ambient gas \citep{1990ApJ...364..555P} and compact H \textsc{ii} regions \citep{1987Natur.326...49B}, and is thought to be related to the formation of complex organic molecules \citep{2009A&A...504..891O}. Since both $^{13}$CH$_3$OH and CH$_3$OH can be detected by CSST/HSTDM in all three cases, one can use them to derive the $^{13}$C/$^{12}$C ratio.

We have considered three assumed cases to give an overview of which molecules can be detected, but they may not be the environments where CH$_3$OH and $^{13}$CH$_3$OH reside in. More realistically, we revisit the spectral lines of $^{13}$CH$_3$OH and CH$_3$OH which reside in Orion-KL with an angular size of 12\,$''$ \citep{Schilke_2001} and different excitation temperatures. Results show that they both can be detected when the excitation temperature is higher than $\sim$12\,K.

\paragraph{CN} The high frequency resolution of CSST/HSTDM enables the detection of CN and its hyper-fine components at $\sim$453\,GHz. The Zeeman effect of CN can be used as a direct measurement of strengths of magnetic fields \citep{1996ApJ...456..217C} and has already been detected toward a few dense clouds through its $N=1\to 0$ transition at $\sim$113\,GHz \citep{1999ApJ...514L.121C}. The CN $N=4\to3$ transition (452$-$454\,GHz) has 19 hyper-fine components, of which six have relatively strong intensities. We adopt an average angular size of 58\,$''$ \citep{2002ApJ...578..211S} for several galactic molecular clouds, and find that these six lines are detectable from $T_{\mathrm{ex}}\simeq$ 8\,K all along to $\sim$300\,K. But according to the project plan of CSST/HSTDM, polarization capability for measuring magnetic field with the Zeeman effect is not implemented.

\subsubsection{Prebiotic molecules}

\begin{figure}[!htb]
	\centering
	\includegraphics[width=14cm]{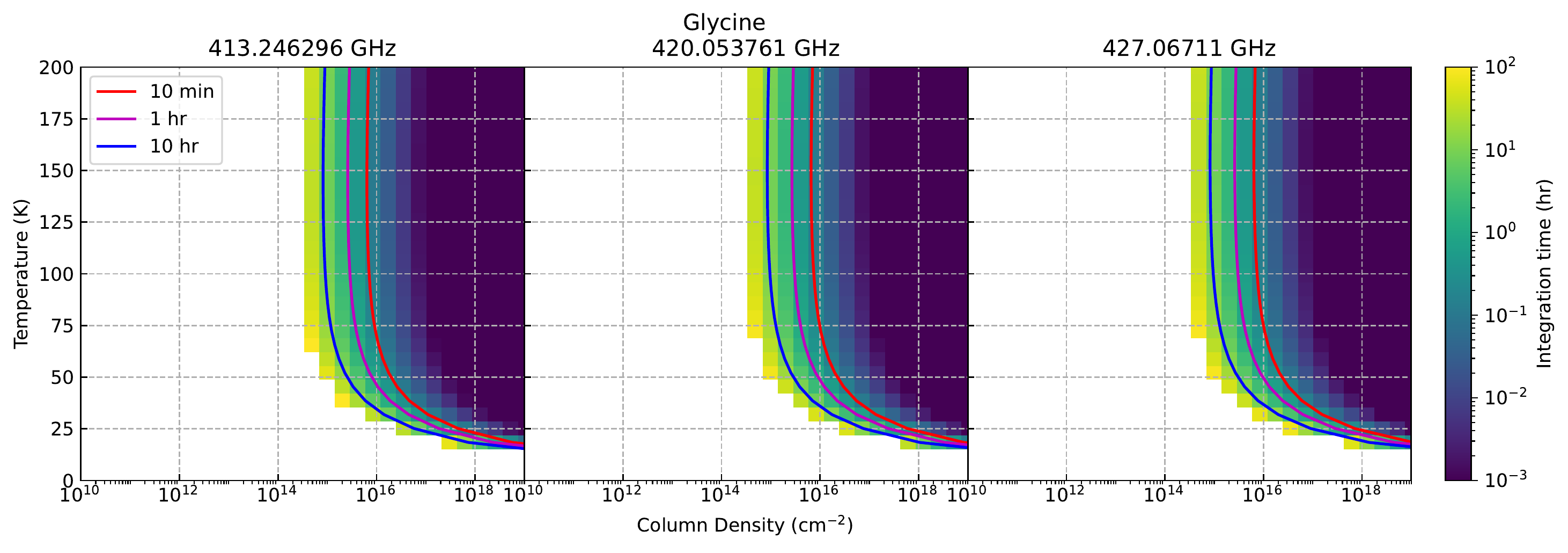}
	\caption{Integration time requirement for CSST/HSTDM to observe three different transitions of glycine. Each pixel in the figure represents a $(T_\text{ex}, N_\text{tot})$ combination, while the color scale indicates the integration time needed for a $3\sigma$ detection, with more than 100 hours set to white.  Red, magenta, and blue contours mark out $(T_\text{ex}, N_\text{tot})$ combinations required for each transition to be detected with an integration time of 10 minutes, 1 hour, and 10 hours, respectively.}
	\label{fig::glycine_obs_CSST}
\end{figure}

Prebiotic molecules are those ``that are thought to be involved in the processes leading to the origin of life'' \citep{2009ARA&A..47..427H}. Amino acids, the basic building blocks of living creatures, are one of the prebiotic molecules most searched for by astronomers. \cite{1970Natur.228..923K} discovered five abundant amino acids in meteorites, and now more than 80 species of amino acids were detected in meteorites (\citealt{1997Sci...275..951C, 2017NatSR...7..636K}, etc.). While no spectral line of interstellar amino acids has been unambiguously confirmed, their possible precursors such as formic acid (HCOOH) \citep{1971ApJ...163L..41Z}, amino acetonitrile (NH$_2$CH$_2$CN) \citep{2008A&A...482..179B}, and isocyanic acid (HNCO) \citep{1972ApJ...177..619S} have been detected. We carried out several simulations on observing prebiotic molecules to investigate how long it would take if we would like to detect them in different astrophysical environments.
\paragraph{Glycine} As the simplest amino acid, glycine (H$_{\mathrm{2}}$NCH$_{\mathrm{2}}$COOH) can form on the analogous of interstellar grains \citep{2002Natur.416..401B}. Several attempts \citep{2007MNRAS.376.1201C, 2007MNRAS.374..579J} had been made to search for its lowest energy conformer I and higher energy conformer II, but failed. In the consideration of an unambiguous identification, we choose three brightest lines of glycine conformer I (hereafter glycine, for convenience) to investigate their detectability under a wider range of $(T_\text{ex}, N_\text{tot})$ combinations. The difference between three cases in Table \ref{tab::common_environment} thus only lies in the source sizes, which behave as the dilution factor of line intensities. To avoid redundance, we only discuss the detectability of glycine under case 3 (hereafter for all prebiotic molecule discussions). For each line we draw a detectability plot in Fig. \ref{fig::glycine_obs_CSST}, where the integration time requirement for detecting glycine under a certain combination of ($T_{\mathrm{ex}}$, $N_{\mathrm{tot}}$) can be figured out.
\citet{2007MNRAS.374..579J} assumed an excitation temperature of $75$\,K and derived an upper limit of $1.4 \times 10^{15}$\,cm$^{-2}$ for the column density for extend ($\eta = 1$) glycine in an eight-hour observation. As a comparison, assuming the same excitation temperature, CSST/HSTDM could possibly detect glycine with a column density of ${\sim}1 \times 10^{15}$\,cm$^{-2}$ in 10 hr even though a smaller beam filling factor ($\eta \simeq0.23$) is adopted. However, we must note that the actual column density of interstellar glycine (if it exists) could be much lower than the value we assumed here.

\begin{figure}[!htb]
	\centering
	\includegraphics[width=14cm]{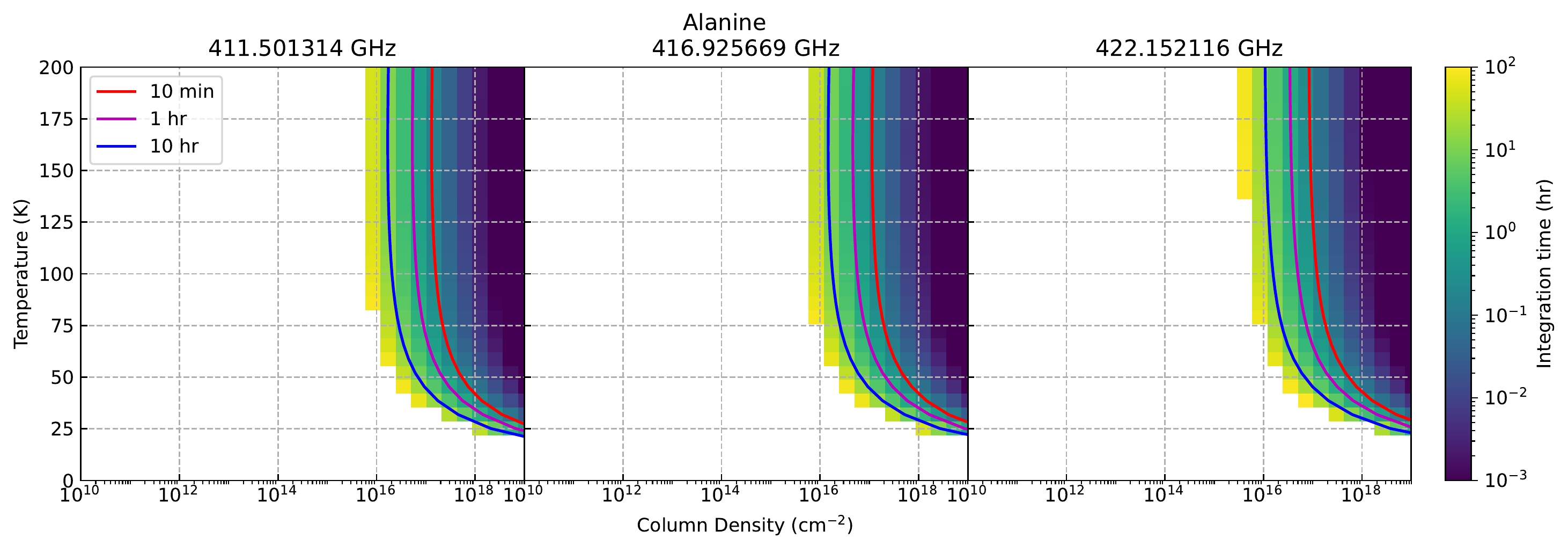}
	\caption{Integration time requirement for CSST/HSTDM to observe three transitions of alanine.}
	\label{fig::alanine_obs_CSST}
\end{figure}

\paragraph{Alanine} Same as glycine, alanine (CH$_3$CHNH$_2$COOH) is also able to form on the analogous of interstellar grains and has two conformers. More importantly, it is one of the only two amino acids (besides glycine) included in CDMS. The detectability of alanine is shown in Fig. \ref{fig::alanine_obs_CSST}.
Compared with those of glycine, the spectral lines of alanine is fainter thus more difficult to be detected with CSST/HSTDM.

\begin{figure}
	\centering
	\includegraphics[width=14cm]{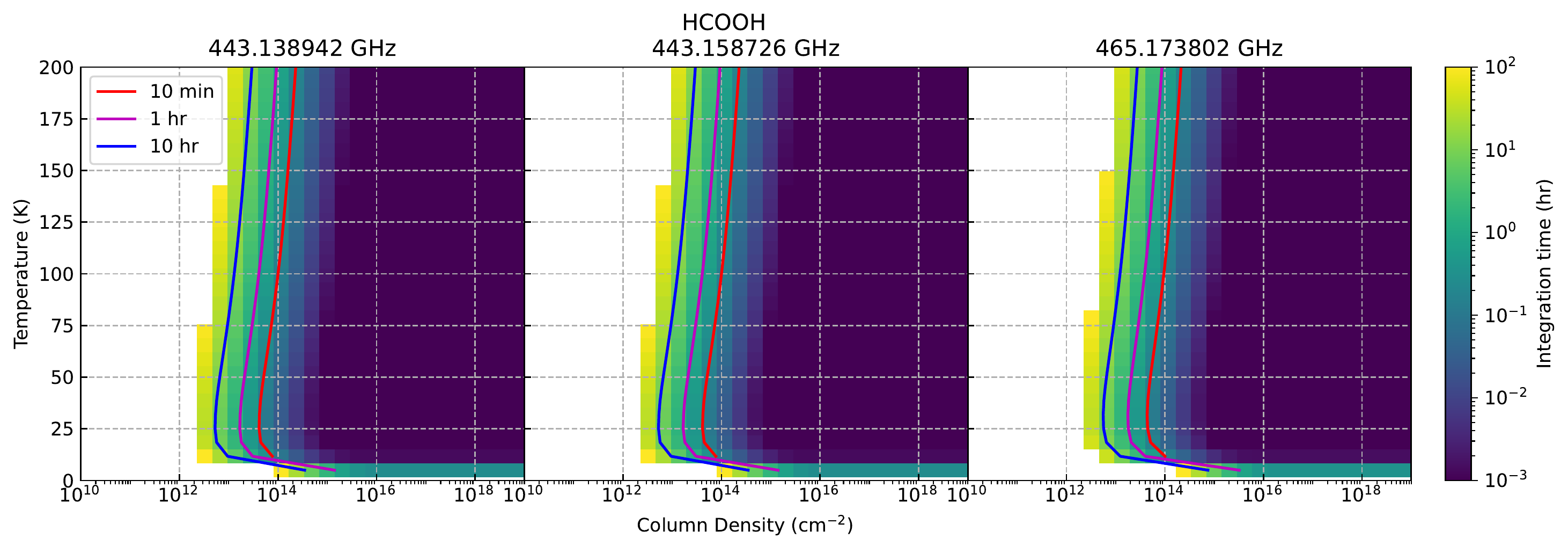}
	\caption{Integration time requirement for CSST/HSTDM to observe three transitions of c-HCOOH.}
	\label{fig::cHCOOH_obs_CSST}
\end{figure}

\paragraph{c-HCOOH}\footnote{The descriptor ``$c$-" represent the two H atoms are on the same side of C-O single bond.} According to \citet{2016A&A...596L...1C}, c-HCOOH is thought to be formed via  fluorescent decay from excited t-HCOOH in ultraviolet irradiated gas, making c-HCOOH not only a related species of amino acids but a tracer of photodissociation regions. Limited by its low column density, this molecule was only detected using the line stacking analysis and $T_{\text{ex}}\simeq21$\,K and $N_{\text{tot}} \simeq 4 \times 10^{12}$\,cm$^{-2}$ was derived \citep{2016A&A...596L...1C}. We can tell from Fig. \ref{fig::cHCOOH_obs_CSST} that CSST/HSTDM could detect c-HCOOH with the same temperature and column density in about 20 hr using regular methods.

As to the line confusion problem, if all previously identified molecules are present as we assumed, lines of these prebiotic molecules may be blended by those of others. More explicitly, among above-mentioned strong lines, the 413.246296\,GHz line of glycine is slightly blended by previously identified molecules; all three c-HCOOH lines are blended, but its several weaker lines around 455 and 509\,GHz are free from confusion.

\subsection{SKA-mid}

The mid-frequency range (0.58$-$25\,GHz) of the initial phase of Square Kilometer Array (hereafter SKA1-mid) will consist of 133 15\,m SKA1 dishes and 64 13.5\,m MeerKAT telescopes\footnote{\url{https://www.skatelescope.org/wp-content/uploads/2021/02/22380\_Construction-Proposal\_DIGITAL\_v3.pdf}}. By means of interferometry, SKA1-mid can reach a high angular resolution ($\eta\simeq$1). Taking advantages of the frequency resolution (13.44\,kHz) and sensitivity (a system temperature of $\sim$33\,K at 10\,GHz), SKA1-mid can be used to search for amino acids and other prebiotic molecules \citep{2004NewAR..48..979C}.

\subsubsection{Molecules that are observable in one hour}

Because of the higher mass and larger sizes, more complex molecules have larger moment of inertia. The same quantized angular momentum corresponds to lower energy levels. Compared with CSST/HSTDM, SKA1-mid covers a lower frequency range, making it more capable of detecting complex molecules. As shown in Table \ref{tab::all_detectable_SKA_1}, \ref{tab::all_detectable_SKA_2}, and \ref{tab::all_detectable_SKA_3}, more complex molecules, especially carbon chain molecules (C$_{\mathrm{n}}$H, HC$_{\mathrm{2n+1}}$N, etc.), are detectable in one hour under all three cases.

\begin{table}[!htb]\tiny
	\begin{center}
		\caption[]{List of potentially detectable molecules with one hour of integration by SKA1-mid in case 1.}\label{tab::all_detectable_SKA_1}		
		\begin{tabular}[5pt]{lllllllll}
			\hline\noalign{\smallskip}
			2 Atoms & 3 Atoms & 4 Atoms & 5 Atoms & 6 Atoms & 7 Atoms & 8 Atoms & 9 Atoms & $>$  9 Atoms \\
			$\mathrm{OH}$	 & $\mathrm{MgNC}$	 & $\mathrm{C_{3}S}$	 & $\mathrm{C_{4}H}$	 & $\mathrm{^{13}\!CH_{3}OH}$	 & $\mathrm{HC_{5}N}$	 & $\mathrm{^{13}\!CH_{2}OHCHO}$ $\!^{*}$	 & $\mathrm{^{13}\!CH_{3}^{13}\!CH_{2}CN}$	 & $\mathrm{C_{2}H_{5}NCO}$ \\
			& $\mathrm{NaCN}$	 & $\mathrm{DNCO}$	 & $\mathrm{HCOCN}$	 & $\mathrm{C_{5}H}$	 & $\mathrm{MgC_{5}N}$	 & $\mathrm{CH_{2}OH^{13}\!CHO}$ $\!^{*}$	 & $\mathrm{^{13}\!CH_{3}CH_{2}^{13}\!CN}$	 & $\mathrm{CH_{3}CH_{2}OCHO}$ \\
			& $\mathrm{SO_{2}}$	 & $\mathrm{HONO}$	 & $\mathrm{MgC_{3}N}$	 & $\mathrm{C_{5}N^{-}}$	 & $c$-$\mathrm{C_{2}H_{4}O}$	 & $\mathrm{CH_{2}OHCHO}$	 & $\mathrm{^{13}\!CH_{3}CH_{2}CN}$	 & $\mathrm{CH_{3}OCH_{2}OH}$ \\
			& $\mathrm{Si_{2}C}$	 & $\mathrm{^{15}\!NH_{3}}$	 & $\mathrm{SiC_{4}}$	 & $\mathrm{CH_{3}OH}$	 & 	 & $\mathrm{H_{2}C_{6}}$	 & $\mathrm{C_{2}H_{5}^{13}\!CN}$	& $aGg'$-$\mathrm{(CH_{2}OH)_{2}}$ \\
			& $\mathrm{TiO_{2}}$	 & $\mathrm{NH_{3}}$	 & 	 & $\mathrm{H_{2}CCNH}$	 & 	 & $\mathrm{H_{2}NCONH_{2}}$	 & $\mathrm{C_{2}H_{5}C^{15}\!N}$	& $gGg'$-$\mathrm{(CH_{2}OH)_{2}}$ \\
			& 	 & 	 & 	 & 	 & 	 & $\mathrm{NH_{2}CH_{2}CN}$	 & $\mathrm{C_{8}H}$	 & $i$-$\mathrm{C_{3}H_{7}CN}$ \\
			& 	 & 	 & 	 & 	 & 	 & 	& $\mathrm{CH_{2}D^{oop}CH_{2}CN}$ $\!^{*}$	 & $n$-$\mathrm{C_{3}H_{7}CN}$ \\
			& 	 & 	 & 	 & 	 & 	 & 	 & $\mathrm{CH_{3}^{13}\!CH_{2}^{13}\!CN}$	 & $s$-$\mathrm{C_{2}H_{5}CHO}$ \\
			& 	 & 	 & 	 & 	 & 	 & 	 & $\mathrm{CH_{3}^{13}\!CH_{2}CN}$	 &  \\
			& 	 & 	 & 	 & 	 & 	 & 	 & $\mathrm{CH_{3}CH_{2}CN}$	 &  \\
			& 	 & 	 & 	 & 	 & 	 & 	 & $\mathrm{CH_{3}CH_{2}OH}$	 &  \\
			& 	 & 	 & 	 & 	 & 	 & 	 & $\mathrm{CH_{3}CH_{2}SH}$	 &  \\
			& 	 & 	 & 	 & 	 & 	 & 	 & $\mathrm{CH_{3}CHDCN}$ $\!^{*}$	 &  \\
			& 	 & 	 & 	 & 	 & 	 & 	 & $\mathrm{CH_{3}OCH_{3}}$	 &  \\
			& 	 & 	 & 	 & 	 & 	 & 	 & $\mathrm{HC_{7}N}$	 &  \\
			\noalign{\smallskip}\hline
		\end{tabular}
		\vspace*{-1.5ex}
		\tablecomments{0.86\textwidth}{The superscript ``$^{\text{oop}}$" in $\mathrm{CH_{2}D^{oop}CH_{2}CN}$ means the D atom are out of the C$_S$ plane. The prefix ``$s$-" of $\mathrm{C_{2}H_{5}CHO}$ represents the formyl and methyl are 0$^{\circ}$ apart from each other.}
	\end{center}
\end{table}
\begin{table}[!htb]\tiny
	\begin{center}
		\caption[]{List of potentially detectable molecules with one hour of integration by SKA1-mid in case 2.}\label{tab::all_detectable_SKA_2}		
		\begin{tabular}[5pt]{llllllll}
			\hline\noalign{\smallskip}
			2 Atoms &  3 Atoms & 4 Atoms & 6 Atoms & 7 Atoms  & 8 Atoms & 9 Atoms & $>$  9 Atoms \\
			\hline\noalign{\smallskip}
			$\mathrm{OH}$	 & $\mathrm{SO_{2}}$	 & $\mathrm{^{15}\!NH_{3}}$	 & $\mathrm{^{13}\!CH_{3}OH}$	 & $\mathrm{HC_{5}N}$	 & $\mathrm{CH_{2}OHCHO}$	 & $\mathrm{^{13}\!CH_{3}^{13}\!CH_{2}CN}$	 & $\mathrm{CH_{3}CH_{2}OCHO}$ \\
			& $\mathrm{TiO_{2}}$	 & $\mathrm{NH_{2}D}$	 & $\mathrm{CH_{3}OH}$	 & 	 & $\mathrm{H_{2}NCONH_{2}}$	 & $\mathrm{^{13}\!CH_{3}CH_{2}^{13}\!CN}$	 & $\mathrm{CH_{3}OCH_{2}OH}$ \\
			& 	 & $\mathrm{NH_{3}}$	 & $\mathrm{H_{2}CCNH}$	 & 	 & $\mathrm{NH_{2}CH_{2}CN}$	 & $\mathrm{^{13}\!CH_{3}CH_{2}CN}$	& $aGg'$-$\mathrm{(CH_{2}OH)_{2}}$ \\
			& 	 & 	 & 	 & 	 & 	 & $\mathrm{C_{2}H_{5}^{13}\!CN}$	& $gGg'$-$\mathrm{(CH_{2}OH)_{2}}$ \\
			& 	 & 	 & 	 & 	 & 	 & $\mathrm{C_{2}H_{5}C^{15}\!N}$	 & $i$-$\mathrm{C_{3}H_{7}CN}$ \\
			& 	 & 	 & 	 & 	 & 	& $\mathrm{CH_{2}D^{oop}CH_{2}CN}$ $\!^{*}$	 & $n$-$\mathrm{C_{3}H_{7}CN}$ \\
			& 	 & 	 & 	 & 	 & 	 & $\mathrm{CH_{3}^{13}\!CH_{2}^{13}\!CN}$	 &  \\
			& 	 & 	 & 	 & 	 & 	 & $\mathrm{CH_{3}^{13}\!CH_{2}CN}$	 &  \\
			& 	 & 	 & 	 & 	 & 	 & $\mathrm{CH_{3}CH_{2}CN}$	 &  \\
			& 	 & 	 & 	 & 	 & 	 & $\mathrm{CH_{3}CHDCN}$ $\!^{*}$	 &  \\
			\noalign{\smallskip}\hline
		\end{tabular}
	\end{center}
\end{table}
\begin{table}\tiny
	\begin{center}
		\caption[]{List of potentially detectable molecules with one hour of integration by SKA1-mid in Case 3.}\label{tab::all_detectable_SKA_3}		
		\begin{tabular}[5pt]{lllllllll}
			\hline\noalign{\smallskip}
			2 Atoms &  3 Atoms & 4 Atoms & 5 Atoms & 6 Atoms & 7 Atoms  & 8 Atoms & 9 Atoms & $>$  9 Atoms \\
			\hline\noalign{\smallskip}
			$\mathrm{OH}$	 & $\mathrm{SO_{2}}$	 & $\mathrm{HONO}$	 & $\mathrm{C_{4}H}$	 & $\mathrm{^{13}\!CH_{3}OH}$	 & $\mathrm{HC_{5}N}$	 & $\mathrm{CH_{2}OHCHO}$	 & $\mathrm{^{13}\!CH_{3}^{13}\!CH_{2}CN}$	 & $\mathrm{CH_{3}CH_{2}OCHO}$ \\
			& $\mathrm{Si_{2}C}$	 & $\mathrm{^{15}\!NH_{3}}$	 & $\mathrm{HCOCN}$	 & $\mathrm{CH_{3}OH}$	 & $c$-$\mathrm{C_{2}H_{4}O}$	 & $\mathrm{H_{2}C_{6}}$	 & $\mathrm{^{13}\!CH_{3}CH_{2}^{13}\!CN}$	 & $\mathrm{CH_{3}OCH_{2}OH}$ \\
			& $\mathrm{TiO_{2}}$	 & $\mathrm{NH_{2}D}$	 & $\mathrm{SiC_{4}}$	 & $\mathrm{H_{2}CCNH}$	 & 	 & $\mathrm{H_{2}NCONH_{2}}$	 & $\mathrm{^{13}\!CH_{3}CH_{2}CN}$	& $aGg'$-$\mathrm{(CH_{2}OH)_{2}}$ \\
			& 	 & $\mathrm{NH_{3}}$	 & 	 & 	 & 	 & $\mathrm{NH_{2}CH_{2}CN}$	 & $\mathrm{C_{2}H_{5}^{13}\!CN}$	& $gGg'$-$\mathrm{(CH_{2}OH)_{2}}$ \\
			& 	 & 	 & 	 & 	 & 	 & 	 & $\mathrm{C_{2}H_{5}C^{15}\!N}$	 & $i$-$\mathrm{C_{3}H_{7}CN}$ \\
			& 	 & 	 & 	 & 	 & 	 & 	& $\mathrm{CH_{2}D^{oop}CH_{2}CN}$ $\!^{*}$	 & $n$-$\mathrm{C_{3}H_{7}CN}$ \\
			& 	 & 	 & 	 & 	 & 	 & 	 & $\mathrm{CH_{3}^{13}\!CH_{2}^{13}\!CN}$	 &  \\
			& 	 & 	 & 	 & 	 & 	 & 	 & $\mathrm{CH_{3}^{13}\!CH_{2}CN}$	 &  \\
			& 	 & 	 & 	 & 	 & 	 & 	 & $\mathrm{CH_{3}CH_{2}CN}$	 &  \\
			& 	 & 	 & 	 & 	 & 	 & 	 & $\mathrm{CH_{3}CH_{2}SH}$	 &  \\
			& 	 & 	 & 	 & 	 & 	 & 	 & $\mathrm{CH_{3}CHDCN}$ $\!^{*}$	 &  \\
			& 	 & 	 & 	 & 	 & 	 & 	 & $\mathrm{CH_{3}OCH_{3}}$	 &  \\
			& 	 & 	 & 	 & 	 & 	 & 	 & $\mathrm{HC_{7}N}$	 &  \\
			\noalign{\smallskip}\hline
		\end{tabular}
	\end{center}
\end{table}

\paragraph{HC$_{\mathrm{2n+1}}$N} Such carbon chain molecules are generally abundant in early stage dense molecular cores before carbon atoms become locked in CO \citep{2013ChRv..113.8981S}. Therefore, they can be used as tracers of dark clouds and starless cores. The case 1 have a similar physical environment with TMC-1, where HC$_5$N, HC$_7$N, HC$_9$N, and HC$_{11}$N have already been detected \citep{1976ApJ...205L.173A, 1978ApJ...223L.105B, 1978ApJ...219L.133K, 2021NatAs...5..188L}. Towards the same source, SKA1-mid can detect HC$_5$N within 5 seconds, HC$_7$N in 10 minutes, and HC$_9$N in 4 hr. \citet{2021NatAs...5..188L} derived an excitation temperature of 6.6\,K and a velocity width of 0.117\,km/s for HC$_{11}$N using line stacking techniques. We use the same parameters as them and find that three brightest HC$_{11}$N lines can be individually detected by SKA1-mid without stacking in 10 minutes.

\subsubsection{Prebiotic molecules}
\label{sec::SKA_prebiotic}

\begin{figure}[!htb]
	\centering
	\includegraphics[width=14cm]{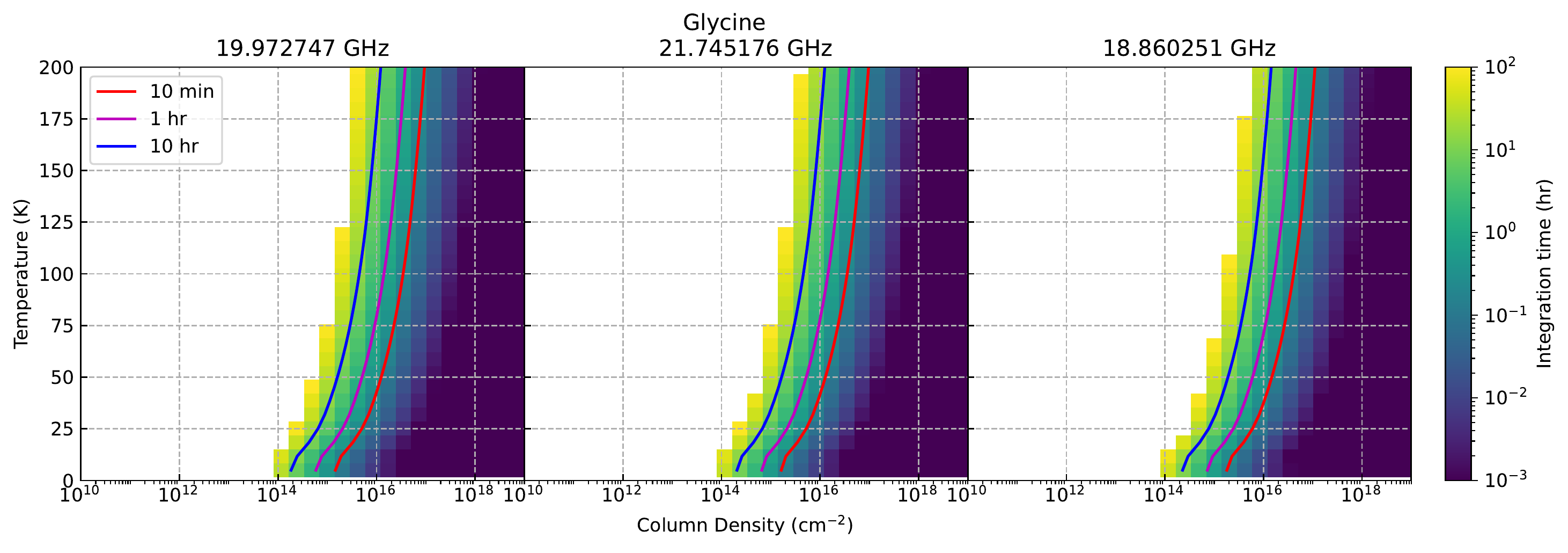}
	\vspace{-4mm}

	\includegraphics[width=14cm]{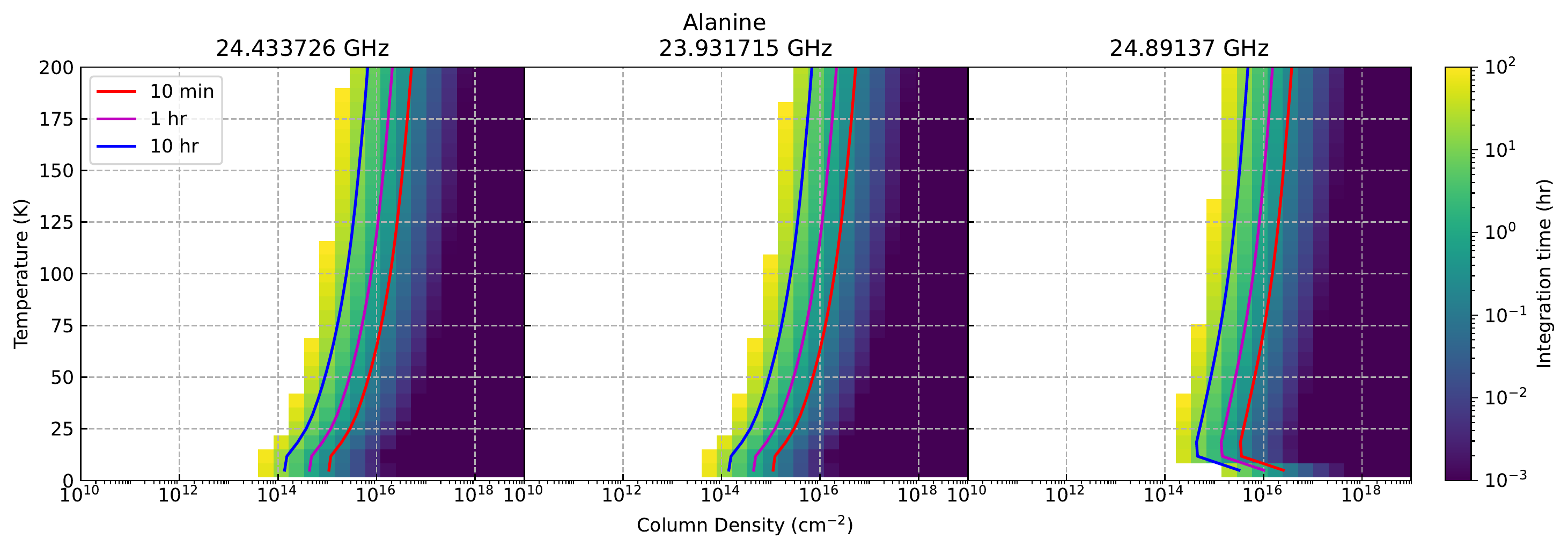}
	\caption{Integration time requirement for SKA1-mid to observe three transitions of glycine (\emph{top}) and alanine (\emph{bottom}).}
	\label{fig::glycine_alanine_obs_SKA}
\end{figure}

\paragraph{Glycine and alanine} In Fig. \ref{fig::glycine_alanine_obs_SKA} we present the detectability of glycine and alanine using SKA1-mid. If the typical temperature of a cold dense molecular cloud is set to $\sim$35\,K \citep{1979MNRAS.186P...5B}, glycine molecules with column densities of ${\sim}1 \times 10^{15}$\,cm$^{-2}$ can be detected in 10 hr. The same holds for alanine.

\begin{figure}[!htb]
	\centering
	\includegraphics[width=14cm]{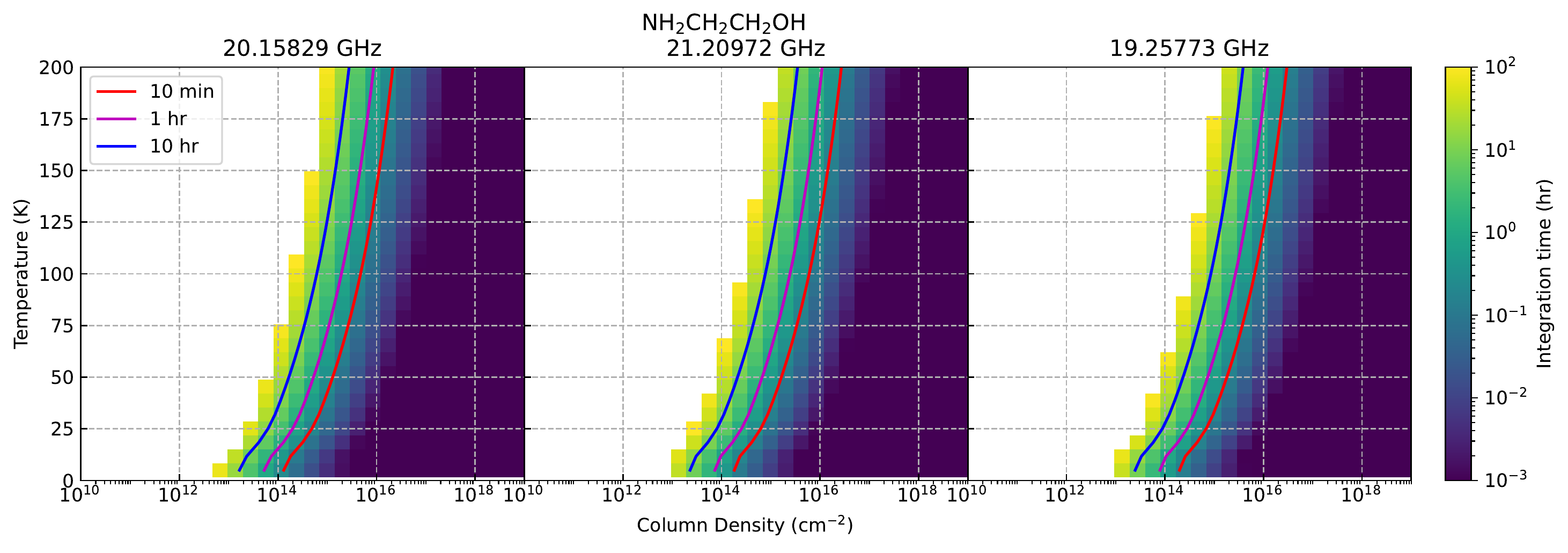}
	\caption{Integration time requirement for SKA1-mid to observe three transitions of NH$_2$CH$_2$CH$_2$OH.}
	\label{fig::NH2CH2CH2OH_SKA}
\end{figure}

\paragraph{Ethanolamine} As the simplest phospholipid head group, ethanolamine (NH$_2$CH$_2$CH$_2$OH) has been identified by  \citet{2021PNAS..11801314R} with $T_{\text{ex}}\simeq 10$\,K and $N_{\mathrm{tot}} \simeq 1.5 \times 10^{13}$\,cm$^{-2}$. For the same excitation temperature and column density SKA1-mid can detect it within 30 hr, as seen in Fig. \ref{fig::NH2CH2CH2OH_SKA}.

\begin{figure}[!htb]
	\centering
	\includegraphics[width=14cm]{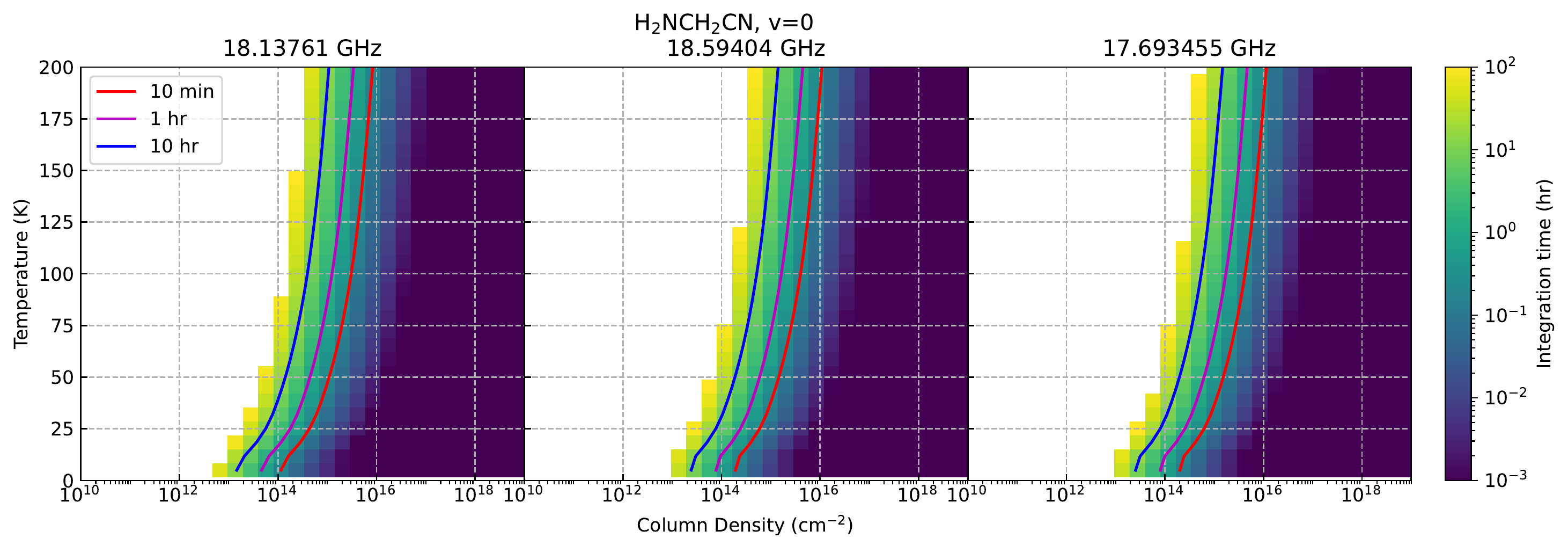}
	\caption{Integration time requirement for SKA1-mid to observe three transitions of H$_2$NCH$_2$CN.}
	\label{fig::H2NCH2CN_SKA}
\end{figure}

\paragraph{Amino acetonitrile} As a possible direct precursor of glycine, amino acetonitrile (H$_2$NCH$_2$CN) is thought to be abundant in ISM \citep{2005GeCoA..69..599P}, and has already been identified by \citet{2008A&A...482..179B} with $T_{\text{ex}}\simeq 100$\,K and $N_{\mathrm{tot}} \simeq 2.8 \times 10^{16}$\,cm$^{-2}$. We present the detectability of H$_2$NCH$_2$CN in Fig. \ref{fig::H2NCH2CN_SKA}. It shows that SKA1-mid can detect H$_2$NCH$_2$CN in 10 hr even if the column density is two orders of magnitude lower.

The working band of SKA1-mid is less crowded than that of CSST/HSTDM, making these prebiotic molecules more possible to be unambiguously identified. More than ten strong lines of glycine and alanine are unblended or only weakly blended by those of previously identified molecules. Although the three strongest amino acetonitrile lines present in Fig. \ref{fig::H2NCH2CN_SKA} are all weakly blended, several weaker lines are free from line confusion.

\subsection{FAST}

Compared with two telescopes mentioned above, the Five-hundred-meter Aperture Spherical radio Telescope (FAST) at present covers a narrower frequency range from 1.05 to 1.45\,GHz, and has a frequency resolution of $\sim$0.48\,kHz. The beamwidth decreases from $\sim$251\,$''$ at 1\,GHz to 140\,$''$ at 1.8\,GHz with a diameter of 300\,m. For simplicity we adopt a constant system temperature of 24\,K \citep{2020RAA....20...64J}, yielding a 3$\sigma$ noise level of $\sim$2.32\,K at 1.4\,GHz in 1 second. Limited by the frequency coverage, only $\mathrm{^{13}\!CH_{3}OH}$ in cases 3 can be detected by FAST within one hour.

If the frequency coverage can slightly extend to a higher range, the OH molecule can be detected by FAST via its four ground state lines: $\mathrm{^{2}\Pi_{3/2}}$ (J=3/2) at 1665.4018 (F=1-1), 1667.3590 (F=2-2), 1612.2310 (F=1-2), and 1720.5300\,MHz (F=2-1). The maser emission of the $\sim$1665 and 1667\,MHz lines are associated with star-forming and their ambient ultracompact H \textsc{ii} regions \citep{1998MNRAS.297..215C}; the 1612\,MHz maser line is used to trace the shell structure of evolved stars \citep{1972A&A....16..204H}; the 1720\,MHz maser line is known to be a probe of shocks \citep{1999ApJ...511..235L}. All these four OH maser lines have been detect by Arecibo telescope \citep{1997ApJS..109..489L, 2018PhRvL.120f1302K}. Using the same facility, \citet{2008ApJ...680..457T} conducted $\sim$500 hr of OH Zeeman observations to derive the magnetic field of nine dark cloud cores, and \citet{2013ApJ...763....8M} measured the Zeeman effect of two main maser lines (1665 and 1669\,MHz). As the largest single-aperture telescope, FAST should be able to detect these four lines as well as their possible maser emission and Zeeman effect.

However, our estimation shows that no prebiotic molecules can be unambiguously detected by FAST within 100 hr  with its current receiver system.

\subsection{Detection by absorption in the presence of a background continuum source}
\begin{figure}[!htb]
	\centering
	\scalebox{0.6}{
	\begin{tikzpicture}
		\draw[line width = 3pt](0, 6)--(0,0);
		\node[rectangle,fill=white,inner sep=0pt] (CMB) at (-0.7,3) {$\text{CMB}$};
		\draw[line width = 1pt, color=cyan](12,3) -- ++(170:9);
		\draw[line width = 1pt, color=cyan](12,3) -- ++(190:9);
		\fill[cyan, opacity = 0.3](3.8,3)ellipse[x radius=0.6,y radius=1.42];
		\draw[cyan, line width=1pt] ([shift=(170:4)]12,3) arc (170:190:4);
		\node[rectangle,fill=white,inner sep=0pt] (fg) at (7.5,3) {\textcolor{cyan}{$\theta_{\text{fg}}$}};
		\node[star, fill=red, inner sep=7.42pt](star) at (1.9,2.948) {\textcolor{red}{a}};
		\draw[line width = 1pt, color=red](12,3) -- ++(176:10.5);
		\draw[line width = 1pt, color=red](12,3) -- ++(184:10.5);
		\draw[red, line width=1pt] ([shift=(176:2.6)]12,3) arc (176:184:2.6);
		\node[rectangle,fill=white,inner sep=0pt] (fg) at (8.8,3) {\textcolor{red}{$\theta_{\text{bg}}$}};
		\draw[line width = 1pt](12,3) -- ++(195:11);
		\draw[line width = 1pt](12,3) -- ++(165:11);
		\draw[line width=1pt] ([shift=(165:6)]12,3) arc (165:195:6);
		\node[rectangle,fill=white,inner sep=0pt] (fg) at (5.7,3) {$\theta_{\text{t}}$};
	\end{tikzpicture}}
	\caption{A sketch of molecular gas (cyan) in front of a background continuum source (red) and CMB (black). Here $\theta_{\text{bg}}$, $\theta_{\text{fg}}$, $\theta_{\text{t}}$ represent the angular sizes of background, foreground, and telescope, respectively.}
	\label{fig::bg_fg_sketch}
\end{figure}
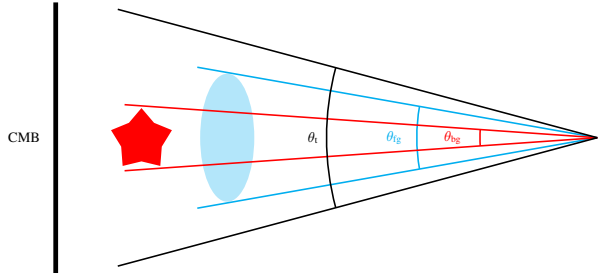

Since many complex organic molecules were detected in massive star-forming regions \citep{2013A&A...559A..47B}, we calculate the spectral lines of prebiotic molecules that are present in front of a bright source, as sketched in Fig. \ref{fig::bg_fg_sketch}. The background source is taken to be a blackbody.

Equation \ref{equ::radiative_e}, taking CMB and filling factors into consideration, can be rewritten as:
\begin{equation}
	I(\nu) = [\eta_{\text{fg}}S_{\nu}(T_{\text{fg}})-\eta_{\text{fg}}S_{\nu}(T_{_{\text{CMB}}}) - \eta_{\text{bg}}B_{\nu}(T_{{\text{bg}}})](1-e^{-\tau_{\nu}}),
\end{equation}
where $T_{\mathrm{bg}}$ represents the temperature of background source, $\eta_{\text{fg}}$ and $\eta_{\text{bg}}$ the filling factors of foreground molecules and the background source, respectively. The $I(\nu)$ here represents the intensity difference between transition lines and the background continuum (i.e. the spectral baseline has been subtracted), one thus may obtain a negative spectral line intensity, i. e. absorption line, if the $\eta_{\text{bg}}B_{\nu}(T_{{\text{bg}}})$ term exceeds the first two terms. Absorption lines with the minimum intensities 3$\sigma$ lower than the local continuum are taken as detectable. Parameters in cases 1, 2, and 3 remain unaltered except for a background source with a temperature of 200\,K and an angular size ($\theta_{\text{bg}}$) of 0.$''$5, which resembles a hot core in Sagittarius B2 molecular cloud \citep{2017A&A...604A..60B}.

The extremely small $\eta_{\text{bg}}$ for CSST/HSTDM ($\sim$10$^{-5}$) and FAST ($\sim$10$^{-6}$) make the spectra almost identical to those with no background source, namely, no absorption feature can be observed with the parameter set assumed here. For SKA1-mid, we continue to assume $\eta_{\text{bg}}=1$ thus a brightness temperature of 200\,K for the background continuum. That is to say, in case 2 of Table \ref{tab::common_environment}, when the excitation temperature of the foreground source is around 200\,K, the small difference between the background and foreground temperature will lead to a low line intensity thus requires a large amount of integration time. For this reason, only $\mathrm{NH_{3}}$, $\mathrm{CH_{3}CH_{2}CN}$, $\mathrm{^{13}\!CH_{3}OH}$, i-$\mathrm{C_{3}H_{7}CN}$, $\mathrm{^{15}\!NH_{3}}$, $\mathrm{OH}$, $\mathrm{C_{2}H_{5}^{13}\!CN}$, and $\mathrm{CH_{3}^{13}\!CH_{2}CN}$ can be detected by SKA1-mid within one hour in this case. Detectable molecules under cases 1 and 3 within one hour using SKA1-mid are listed in Tables \ref{tab::all_detectable_SKA_1_absor} and \ref{tab::all_detectable_SKA_3_absor}, respectively. Many more molecules can be detected in case 1 and 3 when introducing the background source.

\begin{table}[!htb]\tiny
	\begin{center}
		\caption[]{List of potentially detectable molecules in front of a bright source with one hour of integration by SKA1-mid in case 1.}
		\label{tab::all_detectable_SKA_1_absor}		
		\begin{tabular}[5pt]{lllllllll}
			\hline\noalign{\smallskip}
			2 Atoms &  3 Atoms & 4 Atoms & 5 Atoms & 6 Atoms & 7 Atoms  & 8 Atoms & 9 Atoms & $>$  9 Atoms \\
			\hline\noalign{\smallskip}
			$\mathrm{CH}$	 & $\mathrm{C_{2}S}$	 & $\mathrm{C_{3}N}$	 & $\mathrm{C_{4}H}$	 & $\mathrm{^{13}\!CH_{3}OH}$	 & $\mathrm{C_{6}H}$	 & $\mathrm{^{13}\!CH_{2}OHCHO}$ $\!^{*}$	 & $\mathrm{^{13}\!CH_{3}^{13}\!CH_{2}CN}$	 & $\mathrm{C_{2}H_{5}NCO}$ \\
			$\mathrm{KCl}$	 & $\mathrm{CaNC}$	 & $\mathrm{C_{3}S}$	 & $\mathrm{CH_{2}CN}$	 & $\mathrm{C_{5}H}$	 & $\mathrm{C_{6}H^{-}}$	 & $\mathrm{C_{7}H}$	 & $\mathrm{^{13}\!CH_{3}CH_{2}^{13}\!CN}$	 & $\mathrm{C_{2}H_{5}NH_{2}}$ \\
			$\mathrm{OH}$	 & $\mathrm{KNC}$	 & $\mathrm{C_{3}^{34}\!S}$ $\!^{*}$	 & $\mathrm{H_{2}CCO}$	 & $\mathrm{C_{5}N^{-}}$	 & $\mathrm{CH_{2}CHCN}$	 & $\mathrm{CH_{2}ODCHO}$ $\!^{*}$	 & $\mathrm{^{13}\!CH_{3}CH_{2}CN}$	 & $\mathrm{CH_{3}C_{5}N}$ \\
			$\mathrm{SiH}$	 & $\mathrm{MgNC}$	 & $\mathrm{DNCO}$	 & $\mathrm{H_{2}NCN}$	 & $\mathrm{C_{5}S}$	 & $\mathrm{CH_{3}NCO}$, v$_{\mathrm{b}}$=0	 & $\mathrm{CH_{2}OH^{13}\!CHO}$ $\!^{*}$	 & $\mathrm{C_{2}H_{5}^{13}\!CN}$	 & $\mathrm{CH_{3}CH_{2}OCHO}$ \\
			& $\mathrm{NaCN}$	 & $\mathrm{H_{2}CS}$	 & $\mathrm{HCOCN}$	 & $\mathrm{CH_{3}OH}$	 & $\mathrm{H^{13}\!CC_{4}N}$ $\!^{*}$	 & $\mathrm{CH_{2}OHCDO}$ $\!^{*}$	 & $\mathrm{C_{2}H_{5}C^{15}\!N}$	 & $\mathrm{CH_{3}CH_{3}CO}$ \\
			& $\mathrm{SO_{2}}$	 & $\mathrm{HN^{13}\!CO}$ $\!^{*}$	 & $\mathrm{HCCNC}$	 & $\mathrm{CH_{3}SH}$	 & $\mathrm{HC_{2}^{13}\!CC_{2}N}$ $\!^{*}$	 & $\mathrm{CH_{2}CCHCN}$	 & $\mathrm{C_{8}H}$	 & $\mathrm{CH_{3}CHCH_{2}O}$ \\
			& $\mathrm{Si_{2}C}$	 & $\mathrm{HONO}$	 & $\mathrm{HNCCC}$	 & $E$-$\mathrm{HNCHCN}$	 & $\mathrm{HC_{3}^{13}\!CCN}$ $\!^{*}$	 & $\mathrm{CH_{2}CHCHO}$	& $\mathrm{CH_{2}D^{oop}CH_{2}CN}$ $\!^{*}$	 & $\mathrm{CH_{3}OCH_{2}OH}$ \\
			& $\mathrm{SiC_{2}}$	 & $\mathrm{HSCN}$	 & $\mathrm{MgC_{3}N}$	 & $\mathrm{H_{2}CCNH}$	 & $\mathrm{HC_{4}^{13}\!CN}$ $\!^{*}$	 & $\mathrm{CH_{2}OHCHO}$	 & $\mathrm{CH_{3}^{13}\!CH_{2}^{13}\!CN}$	 & $\mathrm{HC_{11}N}$ \\
			& $\mathrm{TiO_{2}}$	 & $\mathrm{^{15}\!NH_{3}}$	 & $\mathrm{SiC_{4}}$	 & $\mathrm{MgC_{4}H}$	 & $\mathrm{HC_{4}NC}$	 & $\mathrm{CH_{3}C_{3}N}$	 & $\mathrm{CH_{3}^{13}\!CH_{2}CN}$	 & $\mathrm{HC_{9}N}$ \\
			& 	 & $\mathrm{NH_{3}}$	 & $c$-$\mathrm{HCOOH}$	 & $\mathrm{SiH_{3}CN}$	 & $\mathrm{HC_{5}N}$	 & $\mathrm{CH_{3}COOH}$, v$_{\mathrm{t}}$=0	 & $\mathrm{CH_{3}C_{4}H}$	 & $\mathrm{NH_{2}CH_{2}CH_{2}OH}$ \\
			& 	 & $c$-$\mathrm{C_{3}H}$	 & $t$-$\mathrm{HCOSH}$	 & $Z$-$\mathrm{HNCHCN}$	 & $\mathrm{HC_{5}^{15}\!N}$ $\!^{*}$	 & $\mathrm{CHDOHCHO}$ $\!^{*}$	 & $\mathrm{CH_{3}CH_{2}CN}$	& $aGg'$-$\mathrm{(CH_{2}OH)_{2}}$ \\
			& 	 & 	 & 	 & $l$-$\mathrm{C_{4}H_{2}}$	 & $\mathrm{HC_{5}O}$	 & $\mathrm{H_{2}C_{6}}$	 & $\mathrm{CH_{3}CH_{2}OH}$	& $gGg'$-$\mathrm{(CH_{2}OH)_{2}}$ \\
			& 	 & 	 & 	 & $l$-$\mathrm{HC_{4}N}$	 & $\mathrm{HC^{13}\!CC_{3}N}$ $\!^{*}$	 & $\mathrm{H_{2}NCONH_{2}}$	 & $\mathrm{CH_{3}CH_{2}SH}$	 & $i$-$\mathrm{C_{3}H_{7}CN}$ \\
			& 	 & 	 & 	 & 	 & $\mathrm{HOCH_{2}CN}$	 & $\mathrm{HC_{5}NH^{+}}$	 & $\mathrm{CH_{3}CHDCN}$ $\!^{*}$	 & $n$-$\mathrm{C_{3}H_{7}CN}$ \\
			& 	 & 	 & 	 & 	 & $\mathrm{MgC_{5}N}$	 & $\mathrm{MgC_{6}H}$	 & $\mathrm{CH_{3}OCH_{3}}$	 & $s$-$\mathrm{C_{2}H_{5}CHO}$ \\
			& 	 & 	 & 	 & 	 & $c$-$\mathrm{C_{2}H_{4}O}$	 & $\mathrm{NH_{2}CH_{2}CN}$	 & $\mathrm{DC_{7}N}$	 &  \\
			& 	 & 	 & 	 & 	 & $s$-$\mathrm{H_{2}CCHOH}$	 & 	 & $\mathrm{HC_{5}^{13}\!CCN}$	 &  \\
			& 	 & 	 & 	 & 	 & 	 & 	 & $\mathrm{HC_{7}N}$	 &  \\
			\noalign{\smallskip}\hline
		\end{tabular}
	\vspace*{-1.5ex}
	\tablecomments{0.86\textwidth}{The prefix ``$l$-" of $\mathrm{C_{4}H_{2}}$ and $\mathrm{HC_{4}N}$ mean the C atoms are arranged in a $linear$ form, while the ``$c$-" in front of $\mathrm{C_{3}H}$ decribe a $cyclic$ form. The ``$t$-" prefix of $\mathrm{HCOSH}$ represents the two H atoms are on the opposite side of the C-S bond. $E$- and $Z$- $\mathrm{HNCHCN}$ describe the H atom and cyanyl on the same and opposite side, respectively.}
	\end{center}
\end{table}

\begin{table}[!htb]\tiny
	\begin{center}
		\caption[]{List of potentially detectable molecules in front of a bright source with one hour of integration by SKA1-mid in case 3.}
		\label{tab::all_detectable_SKA_3_absor}		
		\begin{tabular}[5pt]{lllllllll}
			\hline\noalign{\smallskip}
			2 Atoms &  3 Atoms & 4 Atoms & 5 Atoms & 6 Atoms & 7 Atoms  & 8 Atoms & 9 Atoms & $>$  9 Atoms \\
			\hline\noalign{\smallskip}
			$\mathrm{KCl}$	 & $\mathrm{C_{2}S}$	 & $\mathrm{C_{3}S}$	 & $\mathrm{C_{4}H}$	 & $\mathrm{^{13}\!CH_{3}OH}$	 & $\mathrm{CH_{2}CHCN}$	 & $\mathrm{^{13}\!CH_{2}OHCHO}$ $\!^{*}$	 & $\mathrm{^{13}\!CH_{3}^{13}\!CH_{2}CN}$	 & $\mathrm{C_{2}H_{5}NCO}$ \\
			$\mathrm{OH}$	 & $\mathrm{MgNC}$	 & $\mathrm{DNCO}$	 & $\mathrm{H_{2}CCO}$	 & $\mathrm{C_{5}H}$	 & $\mathrm{H^{13}\!CC_{4}N}$ $\!^{*}$	 & $\mathrm{CH_{2}OH^{13}\!CHO}$ $\!^{*}$	 & $\mathrm{^{13}\!CH_{3}CH_{2}^{13}\!CN}$	 & $\mathrm{CH_{3}CH_{2}OCHO}$ \\
			& $\mathrm{NaCN}$	 & $\mathrm{H_{2}CO}$	 & $\mathrm{H_{2}CNH}$	 & $\mathrm{C_{5}N^{-}}$	 & $\mathrm{HC_{2}^{13}\!CC_{2}N}$ $\!^{*}$	 & $\mathrm{CH_{2}OHCHO}$	 & $\mathrm{^{13}\!CH_{3}CH_{2}CN}$	 & $\mathrm{CH_{3}CH_{3}CO}$ \\
			& $\mathrm{SO_{2}}$	 & $\mathrm{HONO}$	 & $\mathrm{H_{2}NCN}$	 & $\mathrm{C_{5}S}$	 & $\mathrm{HC_{3}^{13}\!CCN}$ $\!^{*}$	 & $\mathrm{CH_{3}C_{3}N}$	 & $\mathrm{C_{2}H_{5}^{13}\!CN}$	 & $\mathrm{CH_{3}OCH_{2}OH}$ \\
			& $\mathrm{Si_{2}C}$	 & $\mathrm{HSCN}$	 & $\mathrm{HCOCN}$	 & $\mathrm{CH_{3}OH}$	 & $\mathrm{HC_{4}^{13}\!CN}$ $\!^{*}$	 & $\mathrm{CH_{3}COOH}$, v$_{\mathrm{t}}$=0	 & $\mathrm{C_{2}H_{5}C^{15}\!N}$	 & $\mathrm{HC_{9}N}$ \\
			& $\mathrm{SiC_{2}}$	 & $\mathrm{^{15}\!NH_{3}}$	 & $\mathrm{MgC_{3}N}$	 & $\mathrm{H_{2}CCNH}$	 & $\mathrm{HC_{5}N}$	 & $\mathrm{H_{2}C_{6}}$	 & $\mathrm{C_{8}H}$	& $aGg'$-$\mathrm{(CH_{2}OH)_{2}}$ \\
			& $\mathrm{TiO_{2}}$	 & $\mathrm{NH_{2}D}$	 & $\mathrm{SiC_{4}}$	 & $\mathrm{MgC_{4}H}$	 & $\mathrm{HC^{13}\!CC_{3}N}$ $\!^{*}$	 & $\mathrm{H_{2}NCONH_{2}}$	& $\mathrm{CH_{2}D^{oop}CH_{2}CN}$ $\!^{*}$	& $gGg'$-$\mathrm{(CH_{2}OH)_{2}}$ \\
			& 	 & $\mathrm{NH_{3}}$	 & 	 & $Z$-$\mathrm{HNCHCN}$	 & $\mathrm{MgC_{5}N}$	 & $\mathrm{MgC_{6}H}$	 & $\mathrm{CH_{3}^{13}\!CH_{2}^{13}\!CN}$	 & $i$-$\mathrm{C_{3}H_{7}CN}$ \\
			& 	 & 	 & 	 & $l$-$\mathrm{C_{4}H_{2}}$	 & $c$-$\mathrm{C_{2}H_{4}O}$	 & $\mathrm{NH_{2}CH_{2}CN}$	 & $\mathrm{CH_{3}^{13}\!CH_{2}CN}$	 & $n$-$\mathrm{C_{3}H_{7}CN}$ \\
			& 	 & 	 & 	 & 	 & 	 & 	 & $\mathrm{CH_{3}CH_{2}CN}$	 & $s$-$\mathrm{C_{2}H_{5}CHO}$ \\
			& 	 & 	 & 	 & 	 & 	 & 	 & $\mathrm{CH_{3}CH_{2}OH}$	 &  \\
			& 	 & 	 & 	 & 	 & 	 & 	 & $\mathrm{CH_{3}CH_{2}SH}$	 &  \\
			& 	 & 	 & 	 & 	 & 	 & 	 & $\mathrm{CH_{3}CHDCN}$ $\!^{*}$	 &  \\
			& 	 & 	 & 	 & 	 & 	 & 	 & $\mathrm{CH_{3}OCH_{3}}$	 &  \\
			& 	 & 	 & 	 & 	 & 	 & 	 & $\mathrm{HC_{7}N}$	 &  \\
			\noalign{\smallskip}\hline
		\end{tabular}
	\end{center}
\end{table}

\begin{figure}[!htb]
	\centering
	\includegraphics[width=14cm]{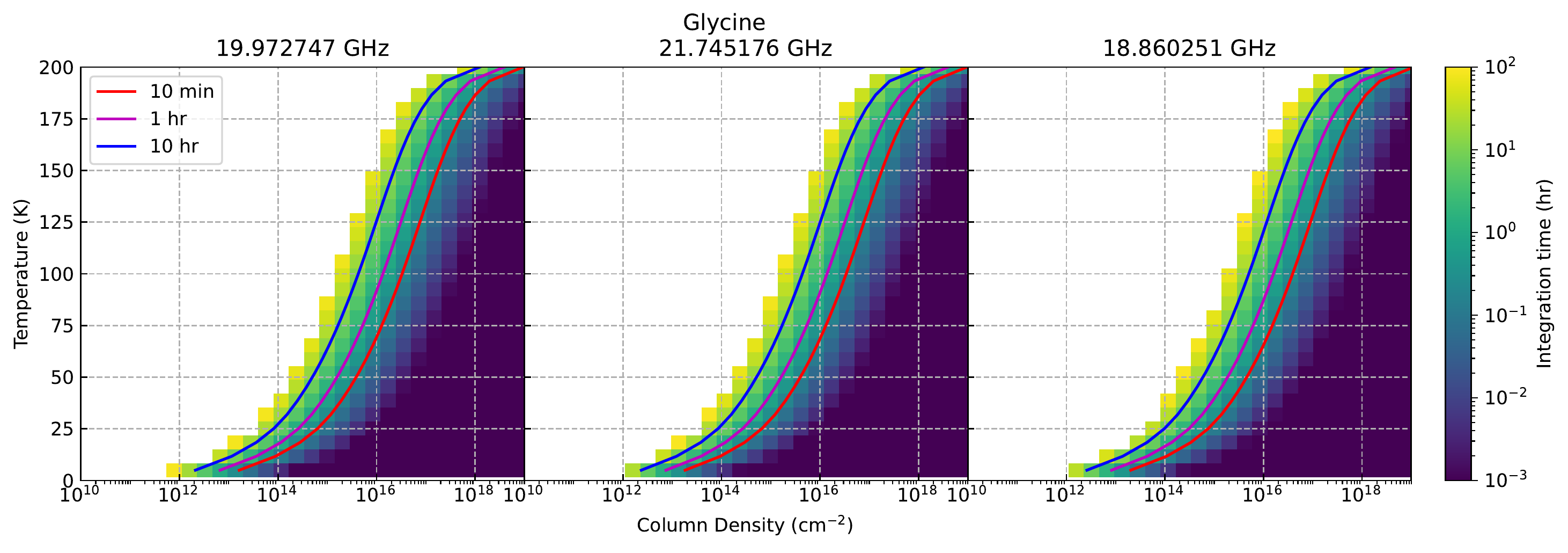}
	\vspace{-4mm}
	
	\includegraphics[width=14cm]{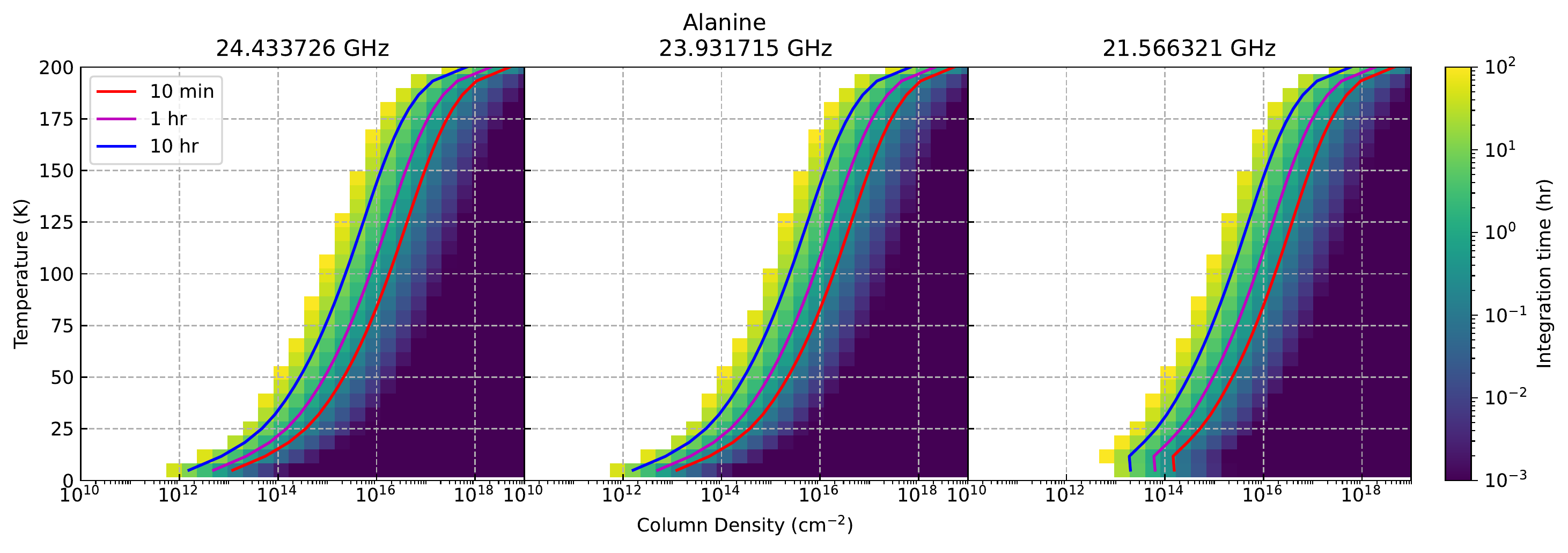}
	\caption{Integration time requirement for SKA1-mid to observe the absorption lines of glycine  (\emph{top}) and alaline (\emph{bottom}) for $T_{\text{bg}}=200$\,K and $\theta_{\text{bg}}=0.5$\,$''$.}
	\label{fig::absorb_gly_ala_SKA}
\end{figure}

Figure \ref{fig::absorb_gly_ala_SKA} shows the detectability of glycine and alanine. We already discussed in Sec. \ref{sec::SKA_prebiotic} that glycine and alanine have the potential to be detected by SAK1-mid via their emission lines in a reasonable time. Under our  assumptions, their absorption lines are even easier to detect when the excitation temperatures are low ($\lesssim$10\,K): it would take 10 hr for SKA1-mid to detect glycine/alanine even if the column density is as low as $7\times10^{12}$/$2\times10^{13}$\,cm$^{-2}$.

\section{Conclusion}
\label{sec::conclusion}

Molecular spectroscopic data are at the base of the molecule identification process. A meta-analysis of related databases gives us a clue about the prospects of future detection of interstellar molecules. The main results of this paper are as follows:
\begin{itemize}
	\item [1)] 
	Most transitions in CDMS and JPL reside in 100$-$1000\,GHz. Transitions within 10$-$1000\,GHz mostly have low frequency uncertainties, and most transitions with high frequency uncertainties are faint (hence unimportant anyway).
	\item [2)]
	Under some common assumptions, the frequency concentration between 100 and 1000\,GHz makes this range more likely to suffer from line confusion problems, causing it more difficult to confirm a molecular species in this frequency range.
	\item [3)]
	More simple molecules can be detected by the HSTDM onboard CSST, while SKA1-mid can detect more complex molecules with one-hour of observation in a variety of ISM environments. Due to the narrow frequency coverage by far, only OH and $\mathrm{^{13}\!CH_{3}OH}$ can possibly be observed by FAST in one hour with plausible assumptions on their column densities.
	\item [4)]
	Prebiotic molecules such as glycine with a column density of ${\sim}1 \times 10^{15}$\,cm$^{-2}$ can be detected by CSST/HSTDM in a 10 hr; SKA1-mid is able to detect glycine and alanine with $N_{\mathrm{tot}}\gtrsim 1 \times 10^{15}$\,cm$^{-2}$ and  $T_{\mathrm{ex}}\simeq 35$\,K in 10 hr. Several strong lines of them are likely not blended by the lines of other molecules hence are ideal for the identification of these molecules.
	\item [5)]
	How a bright background source influences the detectability of molecules depends on its angular size and brightness temperature. For SKA1-mid, the introduction of a bright background source behind a cold dense molecular cloud makes prebiotic molecules easier to detect via absorption features. Compared with emission, glycine/alanine with column densities as low as $7\times10^{12}$/$2\times10^{13}$\,cm$^{-2}$ are able to be detected by SKA1-mid by absorption in 10 hr under our assumptions. Because of the extremely small filling factor, the compact background source has little influence on the detectability of molecules while using CSST/HSTDM and FAST.
	\item [6)]
	When discussing the line confusion problem and detectable molecules in one hour, we simply set the excitation temperature and velocity width of previously identified molecules to the same typical value, respectively, and the column densities of these molecules to their literature or empirical values. These assumptions are oversimplified because the physical parameters may vary dramatically under different interstellar environments. Besides, only one velocity component is considered for each source, which surely underestimate the line confusion problem in crowded regions. When talking about the detectability of prebiotic molecules, the velocity width is fixed to a constant. In the future we will investigate ways to predict the detectability of molecules more realistically.
\end{itemize}

\begin{acknowledgements}
	We thank the referee for his/her constructive suggestions which help to improve the quality of the paper. This work is financially supported by the National Natural Science Foundation of China through grants 12041305 and 11873094, and by the China Manned Space Project.
\end{acknowledgements}
  
\bibliographystyle{raa}
\bibliography{ms2022-0236}

\end{document}